\documentclass[12pt]{iopart}%
\usepackage[T1]{fontenc}%
\usepackage{graphicx}%
\usepackage{lscape}%
\expandafter\let\csname equation*\endcsname=\relax
\expandafter\let\csname endequation*\endcsname=\relax
\usepackage{amsmath}
\usepackage{amssymb}
\usepackage{amsfonts}

\newcommand{\pa}[1]{\left( #1 \right)}
\newcommand{\crochet}[1]{\left[ #1 \right]}
\newcommand{\ac}[1]{\left\{ #1 \right\}}

\newcommand{\s}{\sigma}

\newcommand{\q}{\theta}

\newcommand{\w}{\omega}

\newcommand{\RR}{\mathbb{R}}

\newcommand{\ie}{\textit{i.e.}\ }

\newcommand{\half}{\mbox{$\frac{1}{2}$}}

\newcommand{\del}{\partial}

\newenvironment{aleq}{\begin{equation}\begin{aligned}}{\end{aligned}\end{equation}}
\newenvironment{aleq*}{\begin{equation*}\begin{aligned}}{\end{aligned}\end{equation*}}
\newenvironment{gaeq}{\begin{equation}\begin{gathered}}{\end{gathered}\end{equation}}
\newenvironment{gaeq*}{\begin{equation*}\begin{gathered}}{\end{gathered}\end{equation*}}
\newenvironment{eqe}{\begin{equation}}{\end{equation}}

\begin{document}%
\title[Group analysis of an ideal plasticity model]{Group analysis of an ideal plasticity model}

\author{Vincent Lamothe}

\address{D\'epartement de math\'ematiques et statistiques, Universit\'e de Montr\'eal,\\
C.P. 6128, Succc. Centre-ville, Montr\'eal, (QC) H3C 3J7, Canada}
\ead{lamothe@crm.umontreal.ca}
\begin{abstract}
In this paper, we study the Lie point symmetry group of a system describing an ideal
plastic plane flow in two dimensions in order to find analytical
solutions of the system. The infinitesimal generators that span the Lie algebra
for this system are obtained, six of which are original to this
paper. We completely classify the subalgebras of codimension one and
two into conjugacy classes under the action of the symmetry group.
We apply the symmetry reduction method in order to obtain invariant
and partially invariant solutions. These solutions are of algebraic, trigonometric, inverse trigonometric and elliptic type. Some solutions,
depending on one or two arbitrary functions of one variable, have also been
found. In some cases, the shape of a potentially feasible extrusion
die corresponding to the solution is deduced. These tools could be
used to curve and undulate rectangular rods or slabs, or to shape a ring in an ideal plastic
material.
\end{abstract}
\vspace{2pc}
\submitto{\it{J. Phys.} A}
\maketitle
\section{Introduction}
In this paper, we investigate the plane flow of ideal plastic
materials \cite{Kat:1,Hill,Chak} modelled by the hyperbolic system
of four partial differential equations (PDE) in $q=4$ dependent
variables $\s, \theta, u,v$ and $p=2$ independent variables $x$ and
$y$,
\begin{aleq}\label{eq:1}
&(a)\qquad &&\s_x-2 k \pa{\q_x \cos 2\q+\q_y \sin2\q}=0,\\
&(b)\qquad &&\s_y-2 k \pa{\q_x\sin2\q - \q_y \cos 2\q}=0,\\
&(c)\qquad &&(u_y+v_x)\sin 2\q + (u_x-v_y)\cos2\q=0,\\
&(d)\qquad &&u_x+v_y=0,
\end{aleq}%
where $\s_x=\del \s/\del x$, \textit{etc}. The expressions
(\ref{eq:1}.a), (\ref{eq:1}.b), are the equilibrium equations for
the plane problem. In other words they are the Cauchy differential
equations of motion in a continuous medium where we consider that
the sought quantities do not depend on $z$. These two equations
involve the dependent variables $\s$ and $\q$ that define the stress
tensor; $\s$ is the mean pressure and $\q$ is the angle relative to
the $x$ axis in the counterclockwise direction minus $\pi/4$. The
equation (\ref{eq:1}.c) corresponds, in the plane case, to the
Saint-Venant-Von Mises plasticity theory equations, where $u$ and
$v$ are respectively the velocities in the $x$ axis and $y$ axis
directions. Moreover, we assume that the material is incompressible and hence that the velocity vector is divergenceless. This explains
the presence of the equation (\ref{eq:1}.d) in the considered
system. The positive-definite constant $k$ is called the yield limit
and it is associated with the plastic material. Without loss of
generality we assume that $k=1/2$ (this is the same as re-scaling the
pressure $\sigma$).
\paragraph{}In a recent work \cite{YakhnoYakhno:2010}, the concept of homotopy
of two functions has been used to construct two families of exact
solutions for the system formed by the two first equations in
(\ref{eq:1}). In the papers \cite{Senashov:2007,Senashov:2009}, the
Nada\"{\i} solution \cite{Nadai:circularSol} for a circular cavity
under normal stress and shear and the Prandtl solution
\cite{Prandtl:solPlas} for a bloc compressed between two plates has
been mapped by elements of a symmetry group of the system consisting
of (\ref{eq:1}.a) and (\ref{eq:1}.b), in order to calculate new
solutions. In addition in \cite{Czyz:1}, simple and double Riemann wave
solutions for the system (\ref{eq:1}) were found using the method of
characteristics. However, as is often the case with this method, those
solutions rely on numerical integration for obtaining the velocities
$u$ and $v$. Symmetries of the system (\ref{eq:1}) were found in
\cite{AnninBytevSenashov:1985}. However, the Lie algebra of
symmetries, was not complete because of the absence of the generators
$B_1$, $B_3, B_4,B_4,B_6$ and $K$ (or equivalents) defined by
(\ref{eq:2}) in Section \ref{sec:2} of this work. Moreover, we
found two infinite-dimensional subalgebras spanned by $X_1$ and
$X_2$ provided below in equation (\ref{eq:2b}). The generator $X_1$
was known \cite{AnninBytevSenashov:1985} as a symmetry of the two
first equations in (\ref{eq:1}), but it is shown in this paper to
also be a symmetry of the complete system (\ref{eq:1}). The
generator $X_2$ is a new one. A classification of one-dimensional
subalgebras was performed in \cite{AnninBytevSenashov:1985}. To our
knowledge, no systematic Lie group analysis based on a complete
subalgebra classification in conjugacy classes under the action of
the symmetry group $G$ of the system (\ref{eq:1}) that includes the
new found generators has been done before.
\paragraph*{}The goal of this paper is to systematically investigate the
system (\ref{eq:1}) from the perspective of the Lie group of point
symmetries $G$ in order to obtain analytical solutions. That is, we
obtain in a systematic way all invariant and partially invariant (of
structure defect $\delta=1$ in the sense defined by Ovsiannikov
\cite{Ovsiannikov:Group_Analysis}) solutions under the action of $G$
which are non-equivalent. Invariant solutions are said to be
non-equivalent if they cannot be obtained from one another by a
transformation of $G$ (the solutions are not in the same orbit). In
practice, we apply a procedure developed by J. Patera \textit{et
al}. \cite{PateraWinter:1,PateraWinter:2,WinterPIEEP} that consists
of classifying the subalgebras of $\mathcal{L}$ associated with $G$
into conjugacy classes under the action of $G$. Two subalgebras
$\mathcal{L}_i\subset \mathcal{L}$ and $\mathcal{L}_i'\subset
\mathcal{L}$ are conjugate if $G\mathcal{L}_iG=\mathcal{L}_i'$. For
each conjugacy class, we choose a representative subalgebra, find
its invariants and use them to reduce the initial system
(\ref{eq:1}) to a system in terms of the invariants which involve
fewer variables. We illustrate these theoretical considerations with
some classes of algebraic solutions, some of them in closed form and others defined implicitly. Thereafter, we draw for some solutions and
specific choice of parameters the shape of the corresponding
extrusion die. The applied method relies on the fact that the contours
of the tools must coincide with the flow lines described by the
velocities $u$ and $v$ of the solutions of the problem. For
applications, it is convenient to feed material into the extrusion
die rectilinearly at constant speed. So, the tools illustrated in
this paper were drawn considering this kind of feeding. Based on
mass conservation and on the incompressibility of the materials, we
easily deduce that the curve defining the limit of the plasticity
region for constant feeding speed must obey the ordinary
differential equation (ODE)
\begin{eqe}\label{eq:edoLimPlas}
\frac{dy}{dx}=\frac{V_0-v(x,y)}{U_0-u(x,y)},
\end{eqe}%
where $U_0$, $V_0$ are components of the feeding velocity of the die (or extraction velocity at the
output of the die) respectively along the $x$-axis and $y$-axis. One should note that the
conditions (\ref{eq:edoLimPlas}) are reduced to those required on the limits of the plasticity
region in reference \cite{Czyz:1} when $V_0=0$ and that the curves defining the limits
coincide with slip lines (characteristics), that is when we require
\begin{eqe}\label{eq:char}
dy/dx=\tan\theta(x,y) \text{ or } dy/dx=-\cot\theta(x,y).
\end{eqe}%
Thus the condition (\ref{eq:edoLimPlas}) can be viewed as a relaxation of
the boundary conditions given in \cite{Czyz:1}. The reason we can use these
relaxed conditions is that we choose the contours of the tool to coincide with the flow lines for a
given solution rather than require the flow of material to be parallel to the contours. Using these
relaxed conditions, we can choose (in some limits) the feeding speed and direction for a tool and
this determines the limits of the plasticity region.
\paragraph*{}This paper is organized as follows. In section \ref{sec:2} we give the infinitesimal generators spanning the Lie algebra of symmetries $\mathcal{L}$ for the system (\ref{eq:1}) and the
discrete transformations leaving it invariant. A brief discussion on
the classification of subalgebras of $\mathcal{L}$ in conjugacy
classes follows. Section \ref{sec:3} is concerned with symmetry
reduction. It describes how the symmetry reduction method (SRM) has
been applied to the system (\ref{eq:1}) and the method for finding
partially invariant solutions (PIS). Some examples of solutions are
presented, including invariant and partially invariant ones. We
conclude this paper with a discussion on the obtained results and
 some future outlook.
%######################### section ############# Symétries et classification des sous-algèbres. ###########
\section{Symmetry algebra and classification of its subalgebras}\label{sec:2}
In this section we study the symmetries of the system (\ref{eq:1}).
Following the standard algorithm \cite{Olver:Application_of_Lie},
the Lie symmetry algebra $\mathfrak{L}$ of the system has been
determined. Using the notation $\del_x=\del/\del_x$, \textit{etc}., the Lie algebra of symmetry is spanned by
the fourteen infinitesimal generators
\begin{gaeq}\label{eq:2}
P_1=\del_x,\quad P_2=\del_y,\quad P_3=\del_u,\quad P_4=\del_v,\quad
P_5=\del_\sigma,\\
D_1=x\del_x+y\del_y+u\del_u+v\del_v,\quad
D_2=x\del_x+y\del_y-u\del_u-v\del_v,\\
L=-y\del_x+x\del_y-v\del_u+u\del_v+\del_\theta\\
B_1=-v\del_x+u\del_y,\quad B_2=y\del_u-x\del_v,\\
\begin{aligned}
&B_3=(\sigma+\half\sin2\theta)\del_x-\half\cos2\theta\ \del_y,\quad
&&B_4=-\half\cos2\theta\
\del_x+(\sigma-(1/2)\sin2\theta)\del_y,\\
&B_5=(\sigma-\half\sin2\theta)\del_u+\half\cos2\theta\ \del_v,\quad
&&B_6=\half\cos2\theta\ \del_u+(\sigma+\half \sin2\theta)\del_v,
\end{aligned}\\
\begin{aligned}
K=&(\half x\cos2\theta-y(\sigma+\half
\sin2\theta))\del_x+((\sigma-\half\sin2\theta)x+\half
y\cos2\theta)\del_y\\
&+(\half u\cos2\theta+v(\half
\sin2\theta-\sigma))\del_u+((\sigma+\half \sin2\theta)u-\half
v\cos2\theta)\del_v\\
&+\theta \del_\sigma+\sigma \del_\theta,
\end{aligned}
\end{gaeq}%
and the two generators
\begin{eqe}\label{eq:2b}
X_1=\xi(\sigma,\theta)\del_x+\eta(\sigma,\theta)\del_y,\qquad
X_2=\phi(\sigma,\theta)\del_u+\psi(\sigma,\theta)\del_v,
\end{eqe}%
where the coefficients $\xi$ and $\eta$ must satisfy the two
quasilinear PDEs of the first order
\begin{eqe}\label{eq:2c}
\xi_\sigma=\cos2\theta \xi_\theta+\sin2\theta \eta_\theta,\qquad
\xi_\theta=\cos2\theta \xi_\sigma+\sin2\theta \eta_\sigma,
\end{eqe}%
while the coefficients $\phi$ and $\psi$ must solve the two PDEs, of
the same type as the previous ones,
\begin{eqe}\label{eq:2d}
\phi_\sigma=-\pa{\cos2\theta \phi_\theta+\sin2\theta
\psi_\theta},\qquad \phi_\theta=-\pa{\cos2\theta
\phi_\sigma+\sin2\theta \psi_\sigma}.
\end{eqe}%
Note that $X_1$ and $X_2$ span infinite subalgebras. The generators
$D_1$ and $D_2$ generate dilations in the space of the independent
variables $\ac{x,y}$ and dependent variables $\ac{u,v}$. The
generator $L$ corresponds to a rotation. Moreover, $B_i$,
$i=1,\ldots, 4$ are associated with a type of boost and the $P_i$,
$i=1,\ldots, 5$ generate translations. Of these fourteen generators
eight were already known \cite{AnninBytevSenashov:1985}, but six are
original to this paper. The newly found generators are $B_1,B_3-B_6$
and $K$. In addition, the system (\ref{eq:1}) admits two infinite
dimensional subalgebras. The one spanned by $X_1$ was known
\cite{AnninBytevSenashov:1985} as a symmetry of the two first
equations of the system (\ref{eq:1}), but it still a symmetry of the
entire system (\ref{eq:1}). The infinite subalgebra corresponding to
$X_2$ is original to this paper. The commutation relations for the
$14-$dimensional Lie subalgebra
\begin{eqe}\label{eq:defAlgL}
\mathcal{L}=\ac{B_1,D_2,B_2,L,D_1,B_3,B_4,B_5,B_6,P_1,P_2,P_3,P_4,P_5},
\end{eqe}%
excluding $K$, are shown in table \ref{tab:1}.
\begin{table}[h]
\fontsize{10}{10} \selectfont
\caption{Commutation relations for the algebra $\mathcal{L}$
excluding $K$ in equation (\ref{eq:2}).}\label{tab:1}
\begin{center}
\begin{tabular}{|c| c c c c c c c c c c c c c c|}
  \hline
$\mathcal{L}$ & $B_1$ & $D_2$ & $B_2$ & $D_1$ & $L$ & $P_5$ & $B_3$
& $B_4$ & $B_5$ & $B_6$ & $P_1$ & $P_2$ & $P_3$ & $P_4$\\
\hline $B_1$ & $0$ & $2B_1$ & $-D_2$ & $0$ & $0$ & $0$ & $0$ & $0$ &
$-B_4$ & $B_3$ & $0$ & $0$ & $-P_2$ & $P_1$\\
$D_2$ & $-2B_1$ & $0$ & $2B_2$ & $0$ & $0$ & $0$ & $-B_3$ & $-B_4$ &
$B_5$ & $B_6$ & $-P_1$ & $-P_y$ & $P_3$ & $P_4$\\
$B_2$ & $D_2$ & $-2B_2$ & $0$ & $0$ & $0$ & $0$ & $B_6$ & $-B_5$ &
$0$ & $0$ & $P_4$ & $-P_3$ & $0$ & $0$\\
$D_1$ & $0$ & $0$ & $0$ & $0$ & $0$ & $0$ & $-B_3$ & $-B_4$ & $-B_5$
& $-B_6$ & $-P_1$ & $-P_2$ & $-P_3$ & $-P_4$\\
$L$ & $0$ & $0$ & $0$ & $0$ & $0$ & $0$ & $-B_4$ & $B_3$ & $-B_6$ &
$B_5$ & $-P_2$ & $P_1$ & $-P_4$ & $P_3$\\
$P_5$ & $0$ & $0$ & $0$ & $0$ & $0$ & $0$ & $P_1$ & $P_2$ & $P_3$ &
$P_4$ & $0$ & $0$ & $0$ & $0$\\
$B_3$ & $0$ & $B_3$ & $-B_6$ & $B_3$ & $B_4$ & $-P_1$ & $0$ & $0$ &
$0$ & $0$ & $0$ & $0$ & $0$ & $0$\\
$B_4$ & $0$ & $B_4$ & $B_5$ & $B_4$ & $-B_3$ & $-P_2$ & $0$ & $0$ &
$0$ & $0$ & $0$ & $0$ & $0$ & $0$\\
$B_5$ & $B_4$ & $-B_5$ & $0$ & $B_5$ & $B_6$ & $-P_3$ & $0$ & $0$ &
$0$ & $0$ & $0$ & $0$ & $0$ & $0$\\
$B_6$ & $-B_3$ & $-B_6$ & $0$ & $B_6$ & $-B_5$ & $-P_4$ & $0$ & $0$
& $0$ & $0$ & $0$ & $0$ & $0$ & $0$\\
$P_1$ & $0$ & $P_1$ & $-P_4$ & $P_1$ & $P_2$ & $0$ & $0$ & $0$ & $0$
& $0$ & $0$ & $0$ & $0$ & $0$\\
$P_2$ & $0$ & $P_2$ & $P_3$ & $P_2$ & $-P_1$ & $0$ & $0$ & $0$ & $0$
& $0$ & $0$ & $0$ & $0$ & $0$\\
$P_3$ & $P_2$ & $-P_3$ & $0$ & $P_3$ & $P_4$ & $0$ & $0$ & $0$ & $0$
& $0$ & $0$ & $0$ & $0$ & $0$\\
$P_4$ & $-P_1$ & $-P_4$ & $0$ & $P_4$ & $-P_3$ & $0$ & $0$ & $0$ &
$0$ & $0$ & $0$ & $0$ & $0$ & $0$\\
  \hline
\end{tabular}
\end{center}
\end{table}%
Note that the generators $P_1,P_2,B_3,B_4$ are just particular cases
of $X_1$ in (\ref{eq:2c}) while $P_3,P_4,B_5,B_6$ are particular
cases of $X_2$. Nevertheless, we include them in the classification
under conjugacy classes that we consider in the sequel. We exclude
$K$ from $\mathcal{L}$ because it cannot span a finite Lie algebra
together with $P_1,P_2,P_3,P_4, B_3,B_4,B_5,B_6$. The maximal finite
symmetry Lie algebra that includes $K$ is
\begin{eqe}\label{eq:defAlgS}
\mathcal{S}=\ac{B_1,D_2,B_2,K,L,P_5,D_1}.
\end{eqe}%
The commutation relations for $\mathcal{S}$ are given in table
\ref{tab:2}.
\begin{table}[h]
\fontsize{10}{10} \selectfont
\caption{Commutation relations for the algebra
$\mathcal{S}$.}\label{tab:2}
\begin{center}
\begin{tabular}{|c| c c c c c c c |}
  \hline
$\mathcal{S}$ & $B_1$ & $D_2$ & $B_2$ & $K$ & $L$ & $P_5$  & $D_1$\\
\hline
$B_1$ & $0$ & $2B_1$ & $-D_2$ & $0$ & $0$ & $0$ & $0$\\
$D_2$ & $-2B_1$ & $0$ & $2B_2$ & $0$ & $0$ & $0$ & $0$\\
$B_2$ & $D_2$ & $-2B_2$ & $0$ & $0$ & $0$ & $0$ & $0$\\
$K$ & $0$ & $0$ & $0$ & $0$ & $-P_5$ & $-L$ & $0$\\
$L$ & $0$ & $0$ & $0$ & $P_5$ & $0$ & $0$ & $0$\\
$P_5$ & $0$ & $0$ & $0$ & $L$ & $0$ & $0$ & $0$\\
$D_1$ & $0$ & $0$ & $0$ & $0$ & $0$ & $0$ & $0$\\
  \hline
\end{tabular}
\end{center}
\end{table}%
 One should note that the system
(\ref{eq:1}) is also invariant under the discrete transformations:
\begin{aleq}\label{eq:3}
&R_1:\ x\mapsto -x,\quad &&y\mapsto -y,\quad &&\s\mapsto \s, \quad
&&\q\mapsto \q,\quad  &&u\mapsto u,\quad &&v\mapsto v;\\
&R_2:\ x\mapsto x,\quad &&y\mapsto y,\quad &&\s\mapsto \s, \quad &&\q\mapsto \q,\quad  &&u\mapsto
-u,\quad &&v\mapsto -v.
\end{aleq}%
These transformations $R_1$ and $R_2$ are rotations by an angle
$\pi$ in the plane of independent variables $x,y$ and of dependent
variables $u,v$ respectively that induce the automorphisms of the
Lie algebra $\mathcal{L}$:
\begin{aleq}\label{eq:4}
\mathcal{R}_1:\ &D_1\mapsto D_1,\quad D_2 \mapsto D_2,\quad
B_i\mapsto-B_i,\ i=1..4,\quad B_i\mapsto B_i,\ i=5,6,\\
&P_i\mapsto-P_i,\ i=1,2,\quad P_i\mapsto P_i,\ i=3,4,5,\quad
L\mapsto L,\quad K\mapsto K;\\
\mathcal{R}_2:\ &D_1\mapsto D_1,\quad D_2 \mapsto D_2,\quad
B_i\mapsto-B_i,\ i=1,2,5,6,\quad B_i\mapsto B_i,\ i=3,4,\\
&P_i\mapsto P_i,\ i=1,2,5\quad P_i\mapsto -P_i,\ i=3,4,\quad
L\mapsto L,\quad K\mapsto K.
\end{aleq}%
Since we look for solutions that are invariant and partially invariant
of defect structure $\delta=1$, we only have to classify the
subalgebras of codimension 1 and 2. We consider separately the two
distinct finite dimensional symmetry Lie algebras discussed above.
%########### classification de L #################################
\subsection{\hspace{-2mm} Classification into conjugacy classes of the subalgebra $\mathcal{L}$.}
In order to apply the method of classification into conjugacy
classes developed in \cite{PateraWinter:1,PateraWinter:2}, one first
chooses a decomposition of the Lie algebra. We now describe the
decomposition used for the classification of algebra $\mathcal{L}$.
We begin by factoring $\mathcal{L}$ into the semi-direct sum of the
one-dimensional subalgebra $\ac{P_5}$ and the ideal
$$\mathcal{M}=\ac{B_1,D_2,B_2,L,D_1,B_3,B_4,B_5,B_6,P_1,P_2,P_3,P_4},$$%
\ie
\begin{eqe}\label{eq:5}
\mathcal{L}=\ac{P_5}\rhd \mathcal{M}.
\end{eqe}%
Before we can apply the classification procedure to the semi-direct sum (\ref{eq:5}), we have to know the classification of the subalgebras $\ac{P_5}$ and $M$. The subalgebra $\ac{P_5}$ is a representative of its conjugacy class. In order to classify $M$, consider the following decomposition:
\begin{eqe}\label{eq:6}
\mathcal{M}=\mathcal{F}\rhd \mathcal{N},
\end{eqe}%
where $\mathcal{F}=\ac{B_1,D_2,B_2,L,D_1}$ and $\mathcal{N}=\ac{B_3,B_4,B_5,B_6,P_1,P_2,P_3,P_4}$ is an Abelian subalgebra. The subalgebra $F$ is further decomposed into the direct sum
\begin{eqe}\label{eq:7}
\mathcal{F}=\mathcal{A}\oplus \mathcal{R},
\end{eqe}%
of a simple algebra $\mathcal{A}=\ac{B_1,D_2,B_2}$ and an Abelian
algebra $\mathcal{R}=\ac{L,D_1}$. The classification of the simple
algebra was carried out by Patera and Winternitz in their work
\cite{PateraWinter:1977} on the classification of subalgebras of
real Lie algebras in three and four dimensions. For the
classification of subalgebra $R$, the conjugacy classes are simply
its subspaces since $R$ is Abelian. We then use the Goursat twist to
obtain a list of representative subalgebras for the conjugacy
classes of $F$. The subalgebra $N$ is Abelian, so, once again, all
subspaces are representative subalgebras. Therefore, any further
decomposition may seem superflous. However, since we factored out
the infinitesimal generator $P_5$ and it does not appear in $M$,
subalgebras of type $\mathcal{B}\oplus \ac{0}$ and $\ac{0}\oplus
\mathcal{P}$, where $\mathcal{B}=\ac{B_3,B_4,B_5,B_6}$ and
$\mathcal{P}=\ac{P_1,P_2,P_3,P_4}$, have the same algebraic
structure in $M$, see table \ref{tab:1}. That is, they have the same
commutation relations with elements of $F$. Therefore, taking the
decomposition
$$\mathcal{N}=\mathcal{B}\oplus \mathcal{P},$$
a classification of the splitting subalgebras of the form
$\mathcal{F}_i\rhd\ac{\mathcal{B}\oplus \ac{0}}$, where
$\mathcal{F}_i\subset\mathcal{F}$ is a representative subalgebra of
the classification of $\mathcal{F}$, will lead to that of the
splitting subalgebras of form $\mathcal{F}_i\rhd\ac{\ac{0}\oplus
\mathcal{P}}$ when $(B_3,B_4,B_5,B_6)$ is relabelled
$(P_1,P_2,P_3,P_4)$. Since a splitting subalgebra is a subspace of
form $\mathcal{F}_i\times \mathcal{B}_i$ in which
$\mathcal{B}_i\subset \mathcal{B}$ is an ideal, the splitting
subalgebras with non-zero component in $\mathcal{B}$ and in
$\mathcal{P}$ are classified as follows. For each splitting
subalgebra of form $\mathcal{F}_i\rhd \ac{\mathcal{B}_j\oplus
\ac{0}}$ whose basis vectors $\mathcal{B}_j$ are labeled $b_s$,
$s=1,\ldots, n=\dim\mathcal{B}_j$, one adds to each basis vector
$b_s$ an element $p_s$ of $\mathcal{P}$ under the most general form
$p_s=\sum_{t=1}^4\mu_{st}P_t$, $\mu_{st}\in \RR$. Next, we require
that the space generated by $\ac{b_s+p_s}$, $s=1,\ldots,n$ be an
ideal in $\mathcal{F}_i\times \ac{b_s+p_s}$. This leads to
constraints on the parameters $\mu_{st}$. Once these conditions have
been satisfied, one reduces the range of these parameters as much as
possible by conjugation with $Nor(F_i\rhd\ac{\mathcal{B}_j\oplus
\ac{0}};\exp\mathcal{M})$. This completes the classification of the
splitting subalgebras of $M$. From the list of splitting
subalgebras, we use the procedure described in \cite{PW:1} to find
the subalgebras which are not conjugate to any splitting
subalgebras, the so-called nonsplitting subalgebras. Finally, we
have to classify the semi-direct sum $\ac{P_5}\rhd \mathcal{M}$,
which we do by using the semi-direct sum as developed in
\cite{PateraWinter:1,PateraWinter:2}. Next, we reduce the range of
the parameters as much as possible using the algebra automorphisms
(\ref{eq:4}). The result of the classification of the
one-dimensional subalgebras is presented in table \ref{tab:Ldim1}.
The classification of conjugacy classes for two-dimensional
subalgebras consists of a list of 250 representative subalgebras.
This list can be found in the Appendix.
\begin{table}[h]
\fontsize{10}{10} \selectfont
\begin{center}
\begin{tabular}{|l | l | l |}
\hline
$\mathcal{L}_{1,1}=\ac{B_1}$ &
$\mathcal{L}_{1,19}=\ac{D_2-D_1+B_3+aP_2+\delta P_5}$ &
$\mathcal{L}_{1,37}=\ac{B_3+B_5+\epsilon P_1+ aP_2}$\\
$\mathcal{L}_{1,2}=\ac{B_1+ L+a D_1}$ &
$\mathcal{L}_{1,20}=\ac{D_2-D_1+P_1+aP_5}$ &
$\mathcal{L}_{1,38}=\ac{B_3+B_5+\epsilon P_2}$\\
$\mathcal{L}_{1,3}=\ac{B_1+ L+a D_1+\delta P_5}$ &
$\mathcal{L}_{1,21}=\ac{D_2+D_1+ B_5}$ &
$\mathcal{L}_{1,39}=\ac{B_3+ B_5+\epsilon P_5}$\\
$\mathcal{L}_{1,4}=\ac{B_1+ D_1}$ &
$\mathcal{L}_{1,22}=\ac{D_2+D_1+P_3+aP_5}$ &
$\mathcal{L}_{1,40}=\ac{B_3+ \delta P_2}$\\
$\mathcal{L}_{1,5}=\ac{B_1+ D_1+\delta P_5}$ &
$\mathcal{L}_{1,23}=\ac{D_2+D_1+ B_5 +\delta P_4}$ &
$\mathcal{L}_{1,41}=\ac{B_3+\delta P_2+\epsilon P_4}$\\
$\mathcal{L}_{1,6}=\ac{B_1+ B_5}$ & $\mathcal{L}_{1,24}=\ac{D_2+D_1+
B_5+aP_4+\delta P_5}$ &
$\mathcal{L}_{1,42}=\ac{B_3+\lambda P_3}$\\
$\mathcal{L}_{1,7}=\ac{B_1+ B_5+aP_4+\epsilon P_5}$ &
$\mathcal{L}_{1,25}=\ac{B_1-B_2}$ &
$\mathcal{L}_{1,43}=\ac{B_3+\lambda P_3+a P_4}$\\
$\mathcal{L}_{1,8}=\ac{B_1+ B_5+\epsilon P_4}$ &
$\mathcal{L}_{1,26}=\ac{B_1-B_2+\delta L+a D_1}$ &
$\mathcal{L}_{1,44}=\ac{B_3+\lambda P_4}$\\
$\mathcal{L}_{1,9}=\ac{B_1+ P_3+ P_5}$ &
$\mathcal{L}_{1,27}=\ac{B_1-B_2+\lambda L+a D_1+\delta P_5}$ &
$\mathcal{L}_{1,45}=\ac{B_3+ P_5}$\\
$\mathcal{L}_{1,10}=\ac{B_1+ P_5}$ &
$\mathcal{L}_{1,28}=\ac{B_1-B_2+\lambda D_1}$ &
$\mathcal{L}_{1,46}=\ac{B_5}$\\
$\mathcal{L}_{1,11}=\ac{D_2}$ &
$\mathcal{L}_{1,29}=\ac{B_1-B_2+\lambda D_1+\delta P_5}$ &
$\mathcal{L}_{1,47}=\ac{B_5+P_1+a P_2}$\\
$\mathcal{L}_{1,12}=\ac{D_2+\lambda L +a D_1}$ &
$\mathcal{L}_{1,30}=\ac{B_1-B_2+\lambda P_5}$ &
$\mathcal{L}_{1,48}=\ac{B_5+ P_2+aP_4}$\\
$\mathcal{L}_{1,13}=\ac{D_2+\lambda L+a D_1+\delta P_5}$ &
$\mathcal{L}_{1,31}=\ac{L+a D_1}$ &
$\mathcal{L}_{1,49}=\ac{B_5+ P_5}$\\
$\mathcal{L}_{1,14}=\ac{D_2+\lambda D_1}$ &
$\mathcal{L}_{1,32}=\ac{L+a D_1+\delta P_5}$ &
$\mathcal{L}_{1,50}=\ac{P_1}$\\
$\mathcal{L}_{1,15}=\ac{D_2+\lambda D_1+\delta P_5}$ &
$\mathcal{L}_{1,33}=\ac{D_1}$ &
$\mathcal{L}_{1,51}=\ac{P_1+P_3}$\\
$\mathcal{L}_{1,16}=\ac{D_2+\delta P_5}$ &
$\mathcal{L}_{1,34}=\ac{D_1+\delta P_5}$ &
$\mathcal{L}_{1,52}=\ac{P_3}$\\
$\mathcal{L}_{1,17}=\ac{D_2-D_1+ B_3}$ &
$\mathcal{L}_{1,34}=\ac{B_3}$ & $\mathcal{L}_{1,53}=\ac{P_5}$\\
$\mathcal{L}_{1,18}=\ac{D_2-D_1+B_3+\delta P_2}$ &
$\mathcal{L}_{1,36}=\ac{B_3+B_5}$ &\\
\hline
\end{tabular}
\end{center}
\caption{List of 1-dimensional representative subalgebras of
$\mathcal{L}$. The parameters are $\epsilon=\pm 1$ and
$a,\lambda,\delta,\in\RR$, where $\delta\neq 0$, $\lambda>0$.}
\label{tab:Ldim1}
\end{table}%
%############ classification de S ################################
\subsection{Classification into conjugacy classes of the subalgebra $\mathcal{S}$.}
For the sake of classification, we decompose the seven-dimensional
subalgebra $\mathcal{S}$ into the direct sum
$\mathcal{G}\oplus\ac{D_1}$, with
$\mathcal{G}=\ac{B_1,D_2,B_2,K,L,P_5}$, which is further decomposed
as follows:
$$\mathcal{G}=\mathcal{A}\rhd \mathcal{M},$$%
where $\mathcal{A}$ is the simple algebra of the previous subsection
and $\mathcal{M}=\ac{K,L,P_5}$ is the ideal. Applying the method \cite{PateraWinter:1,PateraWinter:2,WinterPIEEP}, we
proceed to classify all subalgebras of $\mathcal{L}$ into conjugacy classes under the action of the
automorphisms generated by $G$ and the discrete transformations (\ref{eq:3}). In practice, we can
classify the subalgebras under the automorphisms generated by $G$ and decrease the range of the
parameters that appear in the representative subalgebra of a class using the Lie algebra
automorphisms (\ref{eq:4}). The classification results for $\mathcal{S}$ are shown in table \ref{tab:Sdim1} for subalgebras of codimension 1 and in table \ref{tab:Sdim2} for subalgebras
of codimension 2.
\begin{table}[h]
\fontsize{10}{10} \selectfont
\begin{center}
\begin{tabular}{|l | l | l |}
\hline $\mathcal{S}_{1,1}=\ac{B_1}$ &
$\mathcal{S}_{1,14}=\ac{B_1-B_2+\delta L}$ &
$\mathcal{S}_{1,27}=\ac{D_2+\delta_1 L +\delta_2 D_1}$\\
$\mathcal{S}_{1,2}=\ac{B_1+K}$ &
$\mathcal{S}_{1,15}=\ac{B_1-B_2+L+\epsilon P_5}$ &
$\mathcal{S}_{1,28}=\ac{D_2+\delta P_5}$\\
$\mathcal{S}_{1,3}=\ac{B_1+K+\delta D_1}$ &
$\mathcal{S}_{1,16}=\ac{B_1-B_2+L+\epsilon_1 P_5+\epsilon_2 D_1}$ &
$\mathcal{S}_{1,29}=\ac{D_2+\delta_1 L+\delta_2 D_1}$\\
$\mathcal{S}_{1,4}=\ac{B_1+L}$ &
$\mathcal{S}_{1,17}=\ac{B_1-B_2+\delta_1 L+\delta_2 D_1}$ &
$\mathcal{S}_{1,30}=\ac{D_2+\delta D_1}$\\
$\mathcal{S}_{1,5}=\ac{B_1+L+\epsilon P_5}$ &
$\mathcal{S}_{1,18}=\ac{B_1-B_2+\delta P_5}$ &
$\mathcal{S}_{1,31}=\ac{K}$\\
$\mathcal{S}_{1,6}=\ac{B_1+L+\epsilon_1 D_1+\epsilon_2 P_5}$ &
$\mathcal{S}_{1,19}=\ac{B_1-B_2+\delta_1 P_5+\delta_2 D_1}$ &
$\mathcal{S}_{1,32}=\ac{K+\delta D_1}$\\
$\mathcal{S}_{1,7}=\ac{B_1+L+\delta D_1}$ &
$\mathcal{S}_{1,20}=\ac{B_1-B_2+\delta D_1}$ &
$\mathcal{S}_{1,33}=\ac{L}$\\
$\mathcal{S}_{1,8}=\ac{B_1+P_5}$ & $\mathcal{S}_{1,21}=\ac{D_2}$ &
$\mathcal{S}_{1,34}=\ac{L+\delta D_1}$\\
$\mathcal{S}_{1,9}=\ac{B_1+P_5+\delta D_1}$ &
$\mathcal{S}_{1,22}=\ac{D_2+\delta K}$ &
$\mathcal{S}_{1,35}=\ac{L+\epsilon P_5}$\\
$\mathcal{S}_{1,10}=\ac{B_1+D_1}$ &
$\mathcal{S}_{1,23}=\ac{D_2+\delta_1 K+\delta_2 D_1}$ &
$\mathcal{S}_{1,36}=\ac{L+\epsilon_1 P_5+\epsilon_2 D_1}$\\
$\mathcal{S}_{1,11}=\ac{B_1-B_2}$ &
$\mathcal{S}_{1,24}=\ac{D_2+\delta L}$ &
$\mathcal{S}_{1,37}=\ac{P_5}$\\
$\mathcal{S}_{1,12}=\ac{B_1-B_2+\delta K}$ &
$\mathcal{S}_{1,25}=\ac{D_2+L+\epsilon_1 P_5}$ &
$\mathcal{S}_{1,38}=\ac{P_5+\delta D_1}$\\
$\mathcal{S}_{1,13}=\ac{B_1-B_2+\delta_1 K+\delta_2 D_1}$ &
$\mathcal{S}_{1,26}=\ac{D_2+L+\epsilon_1 P_5+\epsilon_2 D_1}$ &
$\mathcal{S}_{1,39}=\ac{D_1}$\\
\hline
\end{tabular}
\end{center}
\caption{List of 1-dimensional representative subalgebras of
$\mathcal{S}$. The parameters are
$\epsilon,\epsilon_1,\epsilon_2=\pm 1$ and
$\delta,\delta_1,\delta_2\in\RR$, $\delta,\delta_1,\delta_2\neq 0$.}\label{tab:Sdim1}
\end{table}
\begin{table}[h]
\fontsize{10}{10} \selectfont
\begin{center}
\begin{tabular}{|l | l |}
\hline
$\mathcal{S}_{2,1}=\ac{B_1,D_2}$ & $\mathcal{S}_{2,32}=\ac{B_1-B_2+\delta K,D_1}$\\
$\mathcal{S}_{2,2}=\ac{B_1,D_2+\delta K}$ & $\mathcal{S}_{2,33}=\ac{B_1-B_2+\delta L,D_1}$\\
$\mathcal{S}_{2,3}=\ac{B_1,D_2+\delta L}$ & $\mathcal{S}_{2,34}=\ac{B_1-B_2+L+\epsilon P_5,D_1}$\\
$\mathcal{S}_{2,4}=\ac{B_1,D_2+\epsilon_1 L+\epsilon_2 P_5}$ & $\mathcal{S}_{2,35}=\ac{B_1-B_2+\delta P_5,D_1}$\\
$\mathcal{S}_{2,5}=\ac{B_1,D_2+\delta P_5}$ & $\mathcal{S}_{2,36}=\ac{B_1-B_2+\delta D_1,K+aD_1}$\\
$\mathcal{S}_{2,6}=\ac{B_1,D_2+\delta_1 K+\delta_2 D_1}$ & $\mathcal{S}_{2,37}=\ac{B_1-B_2+\delta D_1,L+a_1 D_1}$\\
$\mathcal{S}_{2,7}=\ac{B_1,D_2+\epsilon_1 L+\epsilon_2 P_5}$ & $\mathcal{S}_{2,38}=\ac{B_1-B_2+\delta D_1,L+\epsilon_1 P_5+\epsilon_2 D_1}$\\
$\mathcal{S}_{2,8}=\ac{B_1,D_2+\epsilon_1 L +\epsilon_2 P_5+\epsilon_3 D_1}$ & $\mathcal{S}_{2,39}=\ac{B_1-B_2+\delta D_1,P_5+a_1 D_1}$\\
$\mathcal{S}_{2,9}=\ac{B_1,D_2+\delta_1 L+\delta_2 D_1}$ & $\mathcal{S}_{2,40}=\ac{D_2,K}$\\
$\mathcal{S}_{2,10}=\ac{B_1,D_2+\epsilon_1 L+\epsilon_2 P_5+\epsilon_3 D_1}$ & $\mathcal{S}_{2,41}=\ac{D_2,L}$\\
$\mathcal{S}_{2,11}=\ac{B_1,D_2+\delta_1 P_5+\delta_2 D_1}$ & $\mathcal{S}_{2,42}=\ac{D_2,L+\epsilon P_5}$\\
$\mathcal{S}_{2,12}=\ac{B_1,D_2+\delta D_1}$ & $\mathcal{S}_{2,43}=\ac{D_2,P_5}$\\
$\mathcal{S}_{2,13}=\ac{B_1,K}$ & $\mathcal{S}_{2,44}=\ac{D_2,D_1}$\\
$\mathcal{S}_{2,14}=\ac{B_1,L}$ & $\mathcal{S}_{2,45}=\ac{D_2+\delta K,D_1}$\\
$\mathcal{S}_{2,15}=\ac{B_1,L+\epsilon P_5}$ & $\mathcal{S}_{2,46}=\ac{D_2+\delta L,D_1}$\\
$\mathcal{S}_{2,16}=\ac{B_1,P_5}$ & $\mathcal{S}_{2,47}=\ac{D_2+L+\epsilon P_5,D_1}$\\
$\mathcal{S}_{2,17}=\ac{B_1,D_1}$ & $\mathcal{S}_{2,48}=\ac{D_2+\delta P_5,D_1}$\\
$\mathcal{S}_{2,18}=\ac{B_1+K,D_1}$ & $\mathcal{S}_{2,49}=\ac{D_2+\lambda D_1,K+a D_1}$\\
$\mathcal{S}_{2,19}=\ac{B_1+L,D_1}$ & $\mathcal{S}_{2,50}=\ac{D_2+\lambda D_1,L+a D_1}$\\
$\mathcal{S}_{2,20}=\ac{B_1+L+\epsilon P_5,D_2+2\epsilon_2 K}$ & $\mathcal{S}_{2,51}=\ac{D_2+\lambda D_1,L+\epsilon_1 P_5+\epsilon_2 D_1}$\\
$\mathcal{S}_{2,21}=\ac{B_1+L+\epsilon P_5,D_1}$ & $\mathcal{S}_{2,52}=\ac{D_2+\lambda D_1,P_5+a D_1}$\\
$\mathcal{S}_{2,22}=\ac{B_1+P_5,D_1}$ & $\mathcal{S}_{2,53}=\ac{L,P_5+a D_1}$\\
$\mathcal{S}_{2,23}=\ac{B_1+D_1,K+a D_1}$ & $\mathcal{S}_{2,54}=\ac{L+\delta D_1,P_5}$\\
$\mathcal{S}_{2,24}=\ac{B_1+D_1,L+a D_1}$ & $\mathcal{S}_{2,55}=\ac{L+\epsilon_1 D_1,P_5+\epsilon_2 D_1}$\\
$\mathcal{S}_{2,25}=\ac{B_1+D_1,L+\epsilon_1 P_5+\epsilon_2 D_1}$ & $\mathcal{S}_{2,56}=\ac{K,L+\epsilon P_5}$\\
$\mathcal{S}_{2,26}=\ac{B_1+D_1,P_5+aD_1}$ & $\mathcal{S}_{2,57}=\ac{L,P_5}$\\
$\mathcal{S}_{2,27}=\ac{B_1-B_2,K}$ & $\mathcal{S}_{2,58}=\ac{K,D_1}$\\
$\mathcal{S}_{2,28}=\ac{B_1-B_2,L}$ & $\mathcal{S}_{2,59}=\ac{L,D_1}$\\
$\mathcal{S}_{2,29}=\ac{B_1-B_2,L+\epsilon P_5}$ & $\mathcal{S}_{2,60}=\ac{L+\epsilon P_5,D_1}$\\
$\mathcal{S}_{2,30}=\ac{B_1-B_2,P_5}$ & $\mathcal{S}_{2,61}=\ac{P_5,D_1}$\\
$\mathcal{S}_{2,31}=\ac{B_1-B_2,D_1}$ &\\
\hline
\end{tabular}
\end{center}
\caption{List of 2-dimensional representative subalgebras of
$\mathcal{S}$. The parameters are
$\epsilon,\epsilon_1,\epsilon_2=\pm 1$ and
$a,\lambda,\delta,\delta_1,\delta_2\in\RR$,
$\delta,\delta_1,\delta_2\neq 0$, $\lambda>0$.}\label{tab:Sdim2}
\end{table}%
\section{Invariant and partially invariant solutions.}\label{sec:3}
Since equations (\ref{eq:1}.a) and (\ref{eq:1}.b) do not involve the velocity components $u$ and $v$, they can first be solved for $\theta$ and $\sigma$. Next, the result is introduced into the system formed by equations (\ref{eq:1}.c) and (\ref{eq:1}.d) and we look for the solution of this system for the velocity components $u$ and $v$. This system always admits the particular solution
\begin{eqe}\label{eq:soltriv}
u=b_1 y+b_2,\qquad v=-b_1x+b_3,\quad b_1,b_2,b_3\in\RR,
\end{eqe}%
obtained by requiring that the coefficients of the trigonometric functions in (\ref{eq:1}.c) vanish. The velocity field defined by (\ref{eq:soltriv}) forms concentric circles. Therefore, this solution does not establish any relation between the flow velocity and the strain involved in the plastic material. Therefore, it is not an interesting result by itself from the physical point of view. Nevertheless, since the PDEs (\ref{eq:1}.c) and (\ref{eq:1}.d) are linear (assuming that $\theta$ is known), they admit a linear superposition principle. Then, we can add the solution (\ref{eq:soltriv}) to any solution, for the velocity components, of the system (\ref{eq:1}.c), (\ref{eq:1}.d), corresponding to given solutions $\theta$ and $\sigma$ of the system (\ref{eq:1}.a), (\ref{eq:1}.b). Consequently, we can satisfy a much broader family of boundary conditions.
\subsection{Symmetry reduction for the representative subalgebra $B_1$}
\begin{figure}[h]
\begin{center}
\includegraphics[width=8.5cm]{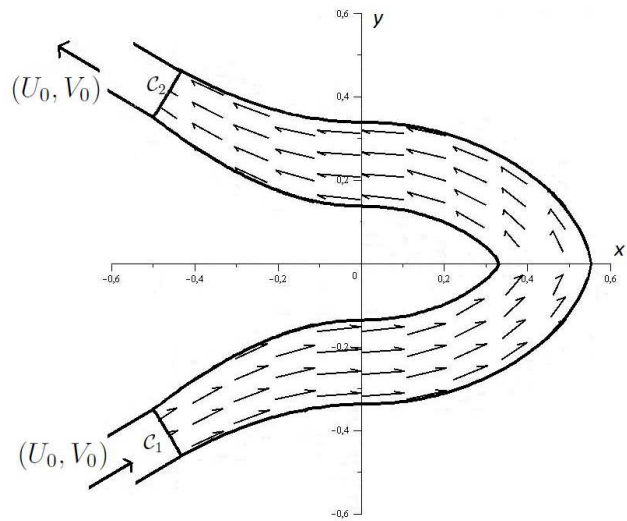}
\end{center}
\caption{Extrusion die corresponding to the solution
(\ref{ex1:eq:6}).} \label{fig:1}
\end{figure}

Consider for illustration the one-dimensional representative subalgebra generated by the infinitesimal generator
\begin{eqe}\label{ex1:eq:1}
B_1=-v\del_x+u\del_y.
\end{eqe}%
Since no derivative with respect to variables $u,v,\theta,\sigma$ appears in $B_1$, it follows that these variables are all invariants of the subalgebra generated by $B_1$. In order to obtain a complete set of functionally independent invariants, one can include also the symmetry variable
\begin{eqe}\label{ex1:eq:2}
\xi=u x - v y.
\end{eqe}%
We look for a solution of the form
\begin{eqe}\label{ex1:eq:3}
u=F(\xi),\quad v=G(\xi),\quad \theta=T(\xi),\quad \sigma=S(\xi),
\end{eqe}%
where $\xi$ is defined by (\ref{ex1:eq:2}). Replacing (\ref{ex1:eq:3}) into the original system (\ref{ex1:eq:1}) and assuming that $1-xF'(\xi)-yG'(\xi)\neq0$, where $F'(\xi)=dF(\xi)/d\xi$, etc., so that we can use the Implicit Function Theorem, we obtain a system of equations with a reduced number of independent variables, where the functions $F,G,T,S$ are to be determined. This system takes the form of four coupled ODEs:
\begin{aleq}\label{ex1:eq:4}
&F(\xi)S'(\xi)-\pa{\cos(2T(\xi))F(\xi)+\sin(2T(\xi))G(\xi)}T'(\xi)=0,\\
&G(\xi)S'(\xi)-\pa{\sin(2T(\xi))F(\xi)-\cos(2T(\xi))G(\xi)}T'(\xi)=0,\\
&\pa{G(\xi)F'(\xi)+F(\xi)G'(\xi)}\sin(2T(\xi))+\pa{F(\xi)F'(\xi)-G(\xi)G'(\xi)}\cos(2T(\xi))=0,\\
&F(\xi)F'(\xi)+G(\xi)G'(\xi)=0.
\end{aleq}%
The solution of this system is
\begin{eqe}\label{ex1:eq:5}
F(\xi)=c_1\cos(T(\xi)),\qquad G(\xi)=c_1\sin(T(\xi)),\quad
S(\xi)=T(\xi)+c_2,
\end{eqe}%
where $T(\xi)$ is an arbitrary function of a single variable and $c_1,c_2$ are constants of integration. The solution is obtained by replacing expressions (\ref{ex1:eq:5}) for $F,G,T,S$ into equation (\ref{ex1:eq:3}). The solution of the original system is given implicitly by relations
\begin{eqe}\label{ex1:eq:6}
u=c_1\cos(ux+yv),\qquad v=c_1\sin(ux+yv),
\end{eqe}%
while the angle $\theta$ and the pressure $\sigma$ are defined by the choice of the arbitrary function $T$ as follows:
\begin{eqe}\label{ex1:eq:7}
\theta=T(ux+yv),\qquad \sigma=T(ux+yv)+c_2.
\end{eqe}%
Since by defining $\theta$ through a certain choice of $T$, we also determine $\sigma$, it follows that equation (\ref{ex1:eq:7}) is a relation defining the pressure $\sigma$ in terms of the angle $\theta$ or vice-versa. Moreover, we can see from equation (\ref{ex1:eq:6}) that the sum of squares of the velocity components $u$ and $v$, is constant. Therefore, since the material is incompressible, the solution preserves the kinetic energy of the plastic material, \ie
$$u^2+v^2=c_1^2.$$%
For the purpose of illustration, consider the function:
\begin{eqe}\label{ex1:eq:8}
T(\xi)=(1/2)\arcsin(\xi),
\end{eqe}%
This particular choice of $T$ allows us to solve relations (\ref{ex1:eq:6}) in order to find the velocities $u$ and $v$ explicitly in terms of $x$ and $y$. The obtained formulas can be expressed in terms of radicals and are very involved. Therefore, they are omitted here. Nevertheless, these formulas can be used to trace the vector fields corresponding to solution (\ref{ex1:eq:6}), where $T$ is defined by (\ref{ex1:eq:8}). An example of such a tool is presented in figure \ref{fig:1} for a flow velocity $c_1=5$. The feeding velocity used is $(U_0,V_0)=(4.30,2.55)$ and the extraction speed is $(U_1,V_1)=(-4.30,2.55)$. The boundaries of the extrusion die are chosen in such a way that they coincide with the flow lines of the velocity field. Therefore, they are solutions of the equation
$$dy/dx=v(x,y)/u(x,y).$$%
For figure 1, the inner boundary corresponds to the initial value $(x_0,y_0)=(-0.5,-0.35)$ and the outer boundary to the initial value $(x_0,y_0)=(-0.43,-0.46)$. The curves $\mathcal{C}_1$ and
$\mathcal{C}_2$ are the limits of the plasticity region with respect to the entrance and exit of the extrusion die. They are solutions of equation (\ref{eq:edoLimPlas}), where, $(U_0,V_0)$ is replaced by $(U_1,V_1)$ for $\mathcal{C}_2$. In order to define the limit of the plasticity region at the ends of the boundary of the tool, the initial data used to trace the curve $\mathcal{C}_1$ are $(x_0,y_0)=(-0.5,-0.35)$ while for $\mathcal{C}_2$ they are $(x_0,y_0)=(-0.5,0.35)$. Numerical integration has been used to identify the boundary of the tool and the limits of the region of plasticity. This type of extrusion die can be used to bend a rectangular rod or a slab of a ideal plastic material. The average pressure and the angle $\theta$, which define the strain tensor inside the tool, are evaluated by formulas (\ref{ex1:eq:7}), where $T$ is defined by (\ref{ex1:eq:8}).
%##########
\subsection{Symmetry reduction for the representative subalgebra $K$}
\begin{figure}[h]
\begin{center}
\includegraphics[width=8cm]{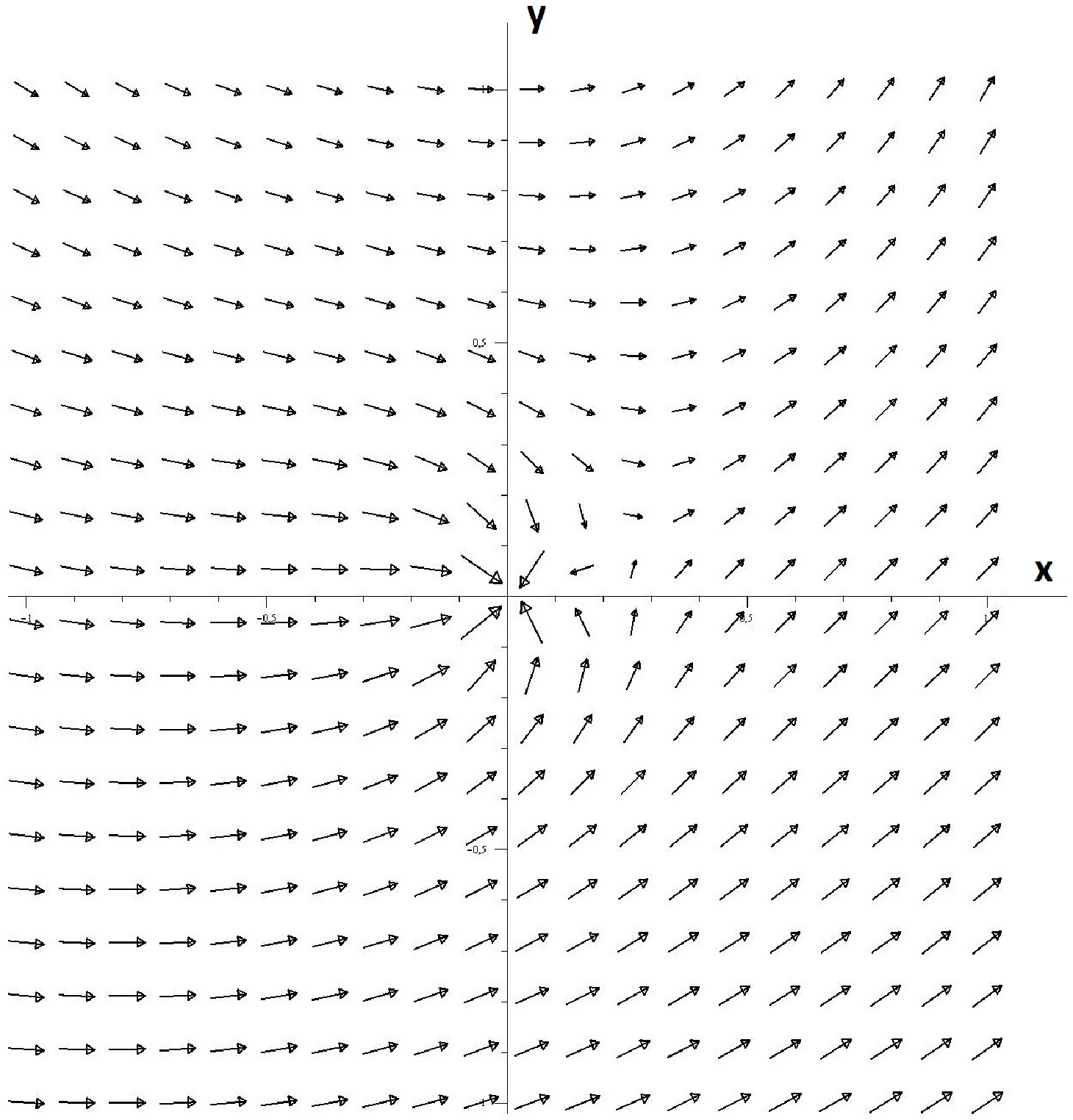}
\end{center}
\caption{Extrusion die corresponding to the solution
(\ref{ex2:eq:7}).} \label{fig:2}
\end{figure}

\begin{figure}[h]
\begin{center}
\includegraphics[width=10cm]{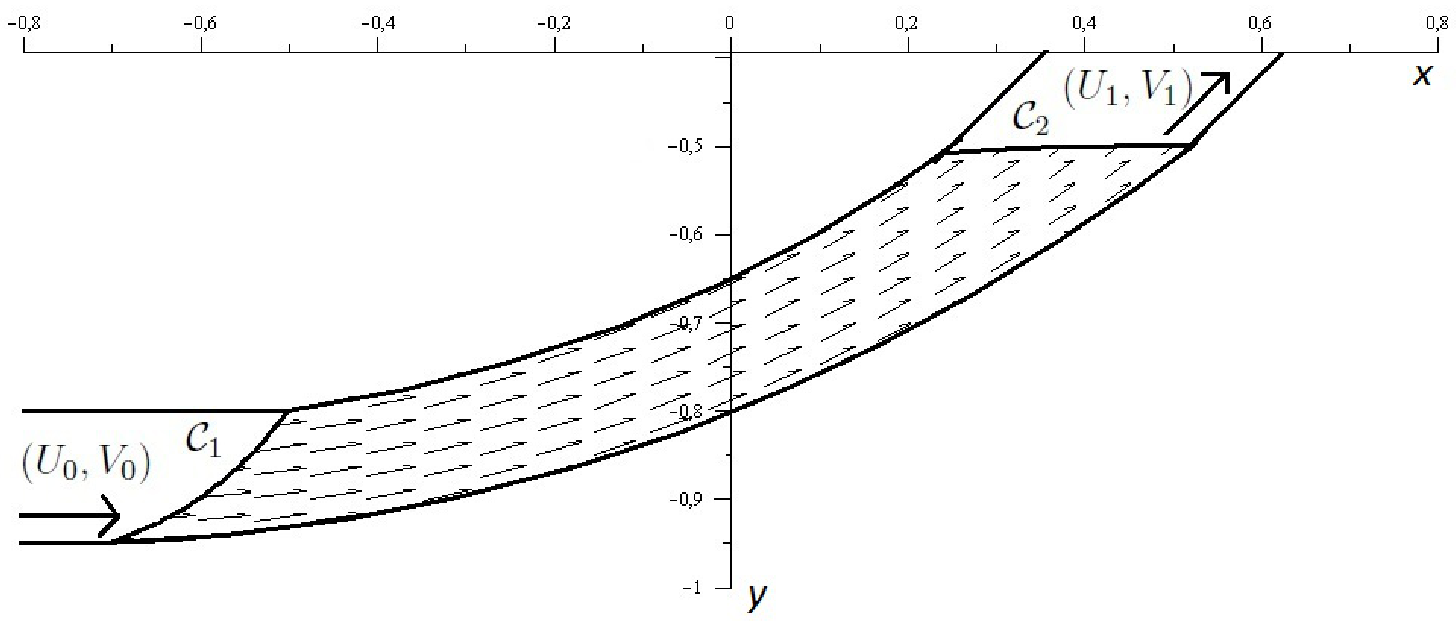}
\end{center}
\caption{Extrusion die corresponding to the solution
(\ref{ex2:eq:7}).} \label{fig:3}
\end{figure}

As an example, we find a partially invariant solution corresponding
to the subalgebra generated by generator $K$ which admits the
following invariants:
\begin{eqe}\label{ex2:eq:1}
\xi=x y \cos(2\theta)-(1/2)(x^2-y^2)\sin(2\theta),\quad F=xu+yv.
\end{eqe}%
and
$$S=\theta^2-\sigma^2,\quad G=u v \cos(2\theta)-(1/2)(u^2-v^2)\sin(2\theta).$$%
In order to obtain a PIS, we use only the two invariants given by
(\ref{ex2:eq:1}). We begin by inverting the first relation in
(\ref{ex2:eq:1}) in order to find $\theta$ as a function of $\xi$.
Next, we introduce the result in the first two equations of system
(\ref{eq:1}). Then, comparing the values of the mixed derivatives of
$\sigma(x,y)$ with respect to $x$ and $y$, we obtain the following PDE for the
quantity $\xi$
\begin{aleq}\label{ex2:eq:2}
&\pa{\xi_{xx}-\xi_{yy}} \pa{((x^2+y^2)^2-4\xi^2)(\xi (x^2-y^2)-x y
\sqrt{(x^2+y^2)^2-4\xi^2})}\\
& -\frac{4xy\xi + (x^2-y^2)\sqrt{(x^2+y^2)^2-4\xi^2}}{x
y\sqrt{(x^2+y^2)^2-4\xi^2}-(x^2-y^2)\xi}\xi_{xy}
+(x^2+y^2)^2\pa{(x+y)\xi_x-(x-y)\xi_y}\\
&\times\pa{(x-y)\xi_x+(x+y)\xi_y}-4(x^2+y^2)^2\xi\pa{x\xi_x+y\xi_y-\xi}=0
\end{aleq}%
There are two particular solutions for $\xi$ to equation (\ref{ex2:eq:2}) defined by
\begin{eqe}\label{ex2:eq:3}
\xi=\frac{1}{2}(x^2+\epsilon y^2),\qquad \epsilon=\pm 1.
\end{eqe}%
Let us consider the case $\epsilon=1$ and introduce this solution into the first relation (\ref{ex2:eq:1}). Solving for $\theta$, we obtain:
\begin{eqe}\label{ex2:eq:4}
\theta=-\frac{1}{2}\arctan\pa{\frac{x^2-y^2}{x y}}.
\end{eqe}%
The mean pressure $\sigma$ is found by quadrature from the first two equations (\ref{eq:1}) in which we have introduced the solution (\ref{ex2:eq:4}) for $\theta$. The result for $\sigma$ is:
\begin{eqe}\label{ex2:eq:5}
\sigma=-(1/2)\ln\pa{x^2+y^2}+c_1,
\end{eqe}
where $c_1$ is a real integration constant.
\paragraph{}Using the form of the second invariant in (\ref{ex2:eq:1}), we look for a solution for the components $u$ and $v$ of the velocity, with the form
\begin{eqe}\label{ex2:eq:6}
u=(y/x)v-F(\xi),
\end{eqe}%
where the symmetry variable $\xi$ is given by (\ref{ex2:eq:3}). By replacing $\theta$ given by
(\ref{ex2:eq:4}) and $u$ by (\ref{ex2:eq:6}) into the system composed of the last two equations in (\ref{eq:1}), then using the compatibility condition of the mixed derivatives of $v$ with respect to $x$ and $y$, we obtain the condition that $F=c_2$, where $c_2$ is a real constant. The solution for $u$ and $v$ is then:
\begin{eqe}\label{ex2:eq:7}
u=\frac{c_2 x}{x^2+y^2}+c_3 y+c_4,\quad v=\frac{c_2 y}{x^2+y^2}-c_3
x+c_5,
\end{eqe}%
where $c_3\ldots, c_5$ are real constants of integration. Note that in the case when $c_4=c_5=0$ and $c_2\neq 0\neq c_3$, the flow lines form logarithmic spirals centered at the origin.
\paragraph{}An example of velocity fields is given in Figure \ref{fig:2} for parameters $c_2=-1$, $c_3=-2$, $c_4=4$ and $c_5=1$ for solution (\ref{ex2:eq:7}). The chosen region, $\crochet{-1,1}\times \crochet{-1,1}$, includes the singularity at the origin. Corresponding to this solution for the same parameters, an extrusion tool is given in Figure \ref{fig:3} for the feeding and extraction velocities $(U_0,V_0)=(5.5,0)$ and
$(U_1,V_1)=(3,3)$ respectively. The curve $\mathcal{C}_1$ is the limit of the plasticity region at the entrance of the extrusion die and $\mathcal{C}_2$ has the same significance at the exit of the extrusion die. The upper contour of the extrusion die is a solution of $dy/dx=v/u$, with $u$ and $v$ defined by (\ref{ex2:eq:7}), for an initial value $y(-0.5)=-0.8$ while, for the lower contour, we have used the initial value $y(-0.7)=-0.95$.
%
%####################### similarity solution ##################################
\subsection{Similarity solution for the angle $\theta$ and corresponding pressure
$\sigma$}\label{sec:3:2} In this section, we find solutions of the system (\ref{eq:1}) for which
the angle $\theta$ is a similarity solution. We propose the solution for $\theta$ in the form
\begin{eqe}\label{eq:formSolProp}
\theta(x,y)=J(\xi(x,y)),
\end{eqe}%
where the symmetry variable is of the form
\begin{eqe}\label{eq:formXiDil}
\xi(x,y)=y/x.
\end{eqe}%
One should note that the solutions, obtained by assuming the hypotheses (\ref{eq:formSolProp}) and (\ref{eq:formXiDil}) on their form, are more general than the invariant solutions corresponding to subalgebras which admit $y/x$ as a symmetry variable. This is so because the invariance requirement, for a given subalgebra, leads to constraints on the form of $\sigma$, which is not the case here. The introduction of
(\ref{eq:formSolProp}), with $\xi$ defined by (\ref{eq:formXiDil}), in the system (\ref{eq:1}.a),
(\ref{eq:1}.b), leads to the system
\begin{aleq}\label{eq:71}
&\s_x(x,y)=2(k/x) \pa{-\xi(x,y)
\cos(J(\xi(x,y)))+\sin(\xi(x,y))} J'(\xi(x,y)),\\
&\s_y(x,y)=-2(k/x)\pa{\xi(x,y)\sin(J(\xi(x,y)))+\cos(J(\xi(x,y)))}J'(\xi(x,y)).
\end{aleq}%
Considering the compatibility condition on mixed derivatives of $\sigma$ relative to $x$ and $y$,
we deduce from (\ref{eq:71}) the following ODE for the function $J$:
\begin{aleq}\label{eq:72}
&\pa{(\xi^2-1)\sin(2J(\xi))+2\xi\cos(2J(\xi))}J''(\xi)+2\pa{\xi\sin(2J(\xi))+\cos(J(\xi))}J'(\xi)\\
&+2\pa{-2\sin(2J(\xi))+(\xi^2-1)\cos(2J(\xi))}(J'(\xi))^2=0,
\end{aleq}%
which has the first integral
\begin{eqe}\label{eq:73}
\pa{(\xi^2-1)\sin(2J(\xi))+2\xi\cos(2J(\xi))}J'(\xi)=c_1,
\end{eqe}%
where $c_1$ is an integration constant. There are two cases to consider to solve the equation
(\ref{eq:73}).
\paragraph*{i).}If $c_1\neq 0$, the solution of (\ref{eq:73}) is given
in implicit form by
\begin{eqe}\label{eq:74}
\frac{\pa{\tan(J(\xi))-\xi}\sqrt{c_1^2-1}}{\pa{\tan(J(\xi))\xi+1}c_1-\xi+\tan(J(\xi))}-\tan\pa{\frac{\sqrt{c_1^2-1}(c_2-J(\xi))}{c_1}}=0,
\end{eqe}%
where $c_2$ is an integration constant. The solution for $\sigma$ is obtained by integrating the
system (\ref{eq:71}) and taking into account the first integral (\ref{eq:73}). We find
\begin{eqe}\label{eq:75}
\s(x,y)=k\pa{\xi(x,y)\cos(2J(\xi(x,y)))-\sin(2J(\xi(x,y)))-2c_1\ln(x)}+c_3,\quad c_3\in \RR
\end{eqe}%
where $\xi(x,y)$ is defined by (\ref{eq:formXiDil}) and $J(\xi)$ is a solution of (\ref{eq:74}). So,
$\theta(x,y)=J(y/x)$ and $\s(x,y)$ are given by (\ref{eq:75}) and are solutions of the system
(\ref{eq:1}.a), (\ref{eq:1}.b), if the function $J(\xi)$ satisfies the algebraic equation
(\ref{eq:74}).
\paragraph*{ii).}If the constant $c_1=0$ in equation (\ref{eq:74}),
then the solution of (\ref{eq:72}) for $J$ is
\begin{eqe}\label{eq:76}
J(\xi)=-\frac{1}{2}\arctan\pa{\frac{2\xi}{\xi^2-1}}.
\end{eqe}%
We subsequently solve (\ref{eq:71}) for $\s(x,y)$ considering (\ref{eq:76}), which leads to the
solution of the system (\ref{eq:1}.a), (\ref{eq:1}.b), given by
\begin{eqe}\label{eq:77}
\theta(x,y)=\frac{1}{2}\arctan\pa{\frac{2xy}{x^2-y^2}},\qquad
\s(x,y)=-2k\arctan\pa{\frac{y}{x}}+c_2.
\end{eqe}%
\subsubsection{Additive separation for the velocities when
$c_1\neq0$.}\label{sec:3:2:1} Already knowing the solution $\theta(x,y)$, $\s(x,y)$, in the case where
$c_1\neq 0$, we still have to compute the solution for $u$ and $v$. One way to proceed is to suppose
that the solution is in the additive separated form
\begin{eqe}\label{eq:78}
u(x,y)=f(x,y)+F(\xi(x,y)),\qquad v(x,y)=g(x,y)+G(\xi(x,y)),
\end{eqe}%
where $\xi(x,y)=y/x$. We introduce (\ref{eq:78}) into the system (\ref{eq:1}.c), (\ref{eq:1}.d),
which gives
\begin{aleq}\label{eq:79}
&\pa{\sin(2J(\xi))-\cos(2J(\xi))}\xi F'(\xi)-\pa{\cos(2J(\xi))+\sin(2J(\xi))}\xi
G'(\xi)\\
&+\pa{\pa{f_y+g_x}\sin(2J(\xi))+\pa{ f_x-g_y}\cos(2J(\xi))}x=0,
\end{aleq}%
\begin{eqe}\label{eq:80}
\pa{f_x+g_y}x+G'(\xi)-\xi F'(\xi)=0,
\end{eqe}%
We must determine which functions $f$ and $g$ will reduce equations (\ref{eq:79}),
(\ref{eq:80}), to a system of ODEs for the single-variable functions $F(\xi)$ and $G(\xi)$. To reach
this goal, we first use as annihilator the infinitesimal generator $(1/2)(D_1+D_2)$ defined by (\ref{eq:2})
that we apply to the equations (\ref{eq:79}), (\ref{eq:80}), to eliminate the presence of the
functions $F$ and $G$. Indeed, the operator $(1/2)(D_1+D_2)$ annihilates any function of $\xi=y/x$. So, we obtain as differential consequences some conditions on the functions $f$ and $g$. We can assume that
$f_x(x,y)+g_y(x,y)\neq0$, otherwise we can show that the only possible solution is the trivial
constant solution for $u$ and $v$. Under this hypothesis, the previous conditions read
\begin{eqe}\label{eq:82}
f_x=-g_y+\zeta_1 (\xi)x^{-1},
\end{eqe}%
\begin{eqe}\label{eq:83}
f_y=-g_x+g_y\zeta_2(\xi)+\zeta_3(\xi)x^{-1},
\end{eqe}%
\begin{eqe}\label{eq:84}
\pa{g_y+(D_1+D_2)\pa{g_y}}\pa{\zeta_2(\xi)\sin(2J(\xi))-2\cos(2J(\xi))}=0,
\end{eqe}%
where the functions of one variable $\zeta_i$, $i=1,2,3$ are arbitrary. Since the left member of
(\ref{eq:84}) is composed of two factors, we must consider two possibilities.
\paragraph*{(a).}We first suppose that
\begin{eqe}\label{eq:85}
g_y+(D_1+D_2)\pa{g_y}=0.
\end{eqe}%
In this case, we find that the functions $f$ and $g$ take the form
\begin{aleq}\label{eq:86}
f(x,y)=&-\int^{\xi(x,y)}\frac{\zeta_1(\xi)-\zeta_4'(\xi)}{\xi}d\xi+\omega_4\ln(y)-\omega_1y+\omega_5,\\
g(x,y)=&\zeta_4(\xi)+\omega_2\ln(x)+\omega_1 x+\omega_3,
\end{aleq}%
where the functions $\zeta_1(\xi), \zeta_4(\xi)$ are arbitrary and the functions $\zeta_2(\xi)$,
$\zeta_3(\xi)$, were chosen to solve the compatibility conditions on mixed derivatives of $f$
relative to $x$ and $y$. We now introduce the solution (\ref{eq:86}) in the system (\ref{eq:82}),
(\ref{eq:83}), which leads to an ODE system for $F$ and $G$, that we omit due to its complexity,
and for which the solutions take the form of quadratures
\begin{aleq}\label{eq:87}
F(\xi)=&\int
\pa{-\zeta'_4(\xi)+\frac{\zeta_1(\xi)}{\xi}+\frac{(\w_4+\w_2\xi)\sin(2J(\xi))}{\xi\pa{(\xi^2-1)\sin(2J(\xi))+2\xi\cos(2J(\xi))}}}d\xi+c_5,\\
G(\xi)=&\int\pa{-\zeta_1(\xi)+\xi F'(\xi)}d\xi+c_4,\qquad c_4,c_5\in \RR.
\end{aleq}%
The last step is to introduce (\ref{eq:86}) and (\ref{eq:87}) in the Ansatz (\ref{eq:78}). Then, the
velocities are
\begin{aleq}\label{eq:88}
u(x,y)=&-\frac{1}{2}\frac{c_5\cos(2J(y/x))}{c_1}+\int^{y/x}\frac{c_6\sin(2J(\xi))J'(\xi)}{c_1\xi}d\xi+c_6\ln(y)-c_4
y+c_7,\\
v(x,y)=&c_5\ln(x)+c_4 x+\int^{y/x}c_1^{-1}
c_5\xi\sin(2J(\xi))J'(\xi)d\xi-(2c_1)^{-1}(c_6\cos(2J(\xi)))+c_8,
\end{aleq}%
where the $c_i$ are integration constants. So, we obtain a solution of the system (\ref{eq:1}) by
defining the angle $\theta$ by (\ref{eq:formSolProp}), (\ref{eq:74}), the mean pressure $\sigma$ by
(\ref{eq:75}) and the velocities $u$ and $v$ by (\ref{eq:88}), with $\xi(x,y)=y/x$.
\paragraph*{(b).}Suppose now that the condition (\ref{eq:84}) is satisfied by requiring
\begin{eqe}\label{eq:89}
\zeta_2(\xi)=2\cot(2J(\xi)).
\end{eqe}%
Then, applying the compatibility condition on mixed derivative of $f$ relative to $x$ and $y$ to
the equations (\ref{eq:82}), (\ref{eq:83}), and considering $\zeta_2$ given by (\ref{eq:89}), we
conclude that the function $g$ must solve the equation
\begin{aleq}\label{eq:90}
&g_{xx}(x,y)+2\cot(2J(\xi))g_{xy}(x,y)-g_{yy}(x,y)+4
x^{-1}\xi\pa{\xi+\cot^2(2J(\xi))}J'(\xi)g_y(x,y)\\
&+x^{-2}\pa{\zeta'_3(\xi)\xi+\zeta'_1(\xi)+\zeta_3(\xi)}=0.
\end{aleq}%
It's a hyperbolic equation everywhere in the domain where $J$ is defined. So, we introduce the
change of variable
\begin{aleq}\label{eq:91}
\phi(x,y)=&x\exp\pa{\int^{\xi(x,y)}\frac{\sin(2J(\xi))}{1+\cos(2J(\xi))+\xi\sin(2J(\xi))}d\xi},\\
\psi(x,y)=&x\exp\pa{\int^{\xi(x,y)}\frac{\sin(2J(\xi))}{-1+\cos(2J(\xi))+\xi\sin(2J(\xi))}d\xi},
\end{aleq}%
which brings the equation (\ref{eq:90}) to the simplified form
\begin{aleq}\label{eq:92}
&g_{\phi,\psi}+\frac{c_1}{2}\pa{\frac{\sin(2J(\phi,\psi))g_\phi}{\psi
(\cos(2J(\phi,\psi)))+1}-\frac{\sin(2J(\phi,\psi))g_\psi}{\phi(\cos(2J(\phi,\psi))-1)}}\\
&-\frac{1}{4}\frac{\sin(2J(\phi,\psi))\xi^2-\sin(2J(\phi,\psi))+2\xi\cos(2J(\phi,\psi))\pa{\zeta_3(\xi)+\zeta'_1(\xi)+\xi\zeta'_3(\xi)}}{J(\phi,\psi)}=0,
\end{aleq}%
where $J(\phi,\psi)$ is defined by
\begin{eqe}\label{eq:37c}
J=(c_1/4)(\psi-\phi).
\end{eqe}%
To solve the equation (\ref{eq:92}) more easily,
we define the function $\zeta_3$ by
\begin{eqe}\label{eq:93}
\zeta_3(\xi)=\xi^{-1}\pa{-\zeta_1(\xi)+J(\xi)+\omega_1}.
\end{eqe}%
So, the solution of (\ref{eq:92}) is
\begin{eqe}\label{eq:94}
g(\phi,\psi)=-(1/2)\pa{\w_1-(1/2)\ln(\psi/\phi)}\cos\pa{c_1\ln(\psi/\psi)}-(1/4)c_1^{-1}\sin\pa{c_1\ln(\psi/\phi)},
\end{eqe}%
which, returning to the initial variables, takes the form
\begin{eqe}\label{eq:95}
g(x,y)=-(1/2)\pa{\w_1-(1/2)c_1^{-1}J(y/x)}\cos(2J(y/x))-(1/4)c_1^{-1}\sin(2J(y/x))+\w_2.
\end{eqe}%
After the introduction of the solution (\ref{eq:95}) for $g$, the function $f$ is given by
quadrature from the equations (\ref{eq:82}), (\ref{eq:83}). The obtained solution for $f$ is
\begin{eqe}\label{eq:96}
f(x,y)=\pa{\int^{\xi(x,y)}\frac{\pa{c_1\w_1-J(\xi)}\sin(2J(\xi))J'(\xi)}{c_1\xi}d\xi}-\int^{\xi(x,y)}\frac{\zeta_1(\xi)}{\xi}d\xi+(c_1+1)\w_1\ln(y)+\w_3.
\end{eqe}%
We now introduce (\ref{eq:95}), (\ref{eq:96}), in (\ref{eq:79}), (\ref{eq:80}), and get $F$ and $G$
by quadrature in the form
\begin{aleq}\label{eq:97}
F(\xi)&=\int^{\xi(x,y)}\frac{\zeta_1(\xi)}{\xi}d\xi+\int^{\xi(x,y)}\frac{(\w_1+J(\xi))\sin(2J(\xi))J'(\xi)}{c_1\xi}d\xi,\\
G(\xi)&=-(1/2)c_1^{-1}\pa{(\w_1+J(\xi))\cos(2J(\xi))+2\sin(2J(\xi))}.
\end{aleq}%
Finally, the substitution of (\ref{eq:95}), (\ref{eq:96}) and (\ref{eq:97}) in (\ref{eq:78})
provides the solution to (\ref{eq:1}.c), (\ref{eq:1}.d):
\begin{aleq}\label{eq:98}
u(x,y)=&(c_1+1)c_4\ln(y)+\int^{y/x}\frac{c_4(c_1+1)\sin(2J(\xi))J'(\xi)}{c_1\xi}d\xi+c_5,\\
v(x,y)=&-\frac{c_4(c_1+1)\cos(2J(y/x))}{c_1}+c_6,
\end{aleq}%
where the $c_i$ are integration constants. So, we have a solution of the system (\ref{eq:1}) by
implicitly defining the angle $\theta$ by (\ref{eq:formSolProp}), (\ref{eq:74}), the mean pressure
by $\sigma$ by (\ref{eq:75}) and the velocities $u$ and $v$ by (\ref{eq:98}), with $\xi(x,y)=y/x$.
%####################################### additive separation for the similarity solution (c_1=0)###################################
\subsubsection{Additive separation for the velocities $u$ and $v$ when
$c_1=0$.}\label{sec:3:2:2}%
\begin{figure}[h]
\begin{center}
\includegraphics[width=10cm]{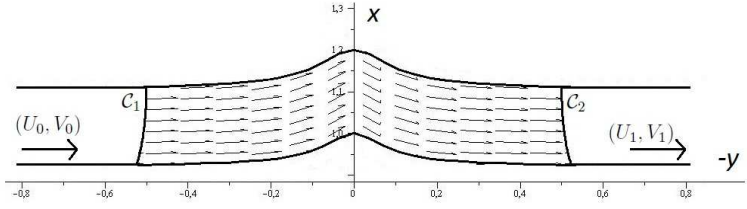}
\end{center}
\caption{Extrusion die corresponding to the solution (\ref{eq:77}), (\ref{eq:101}).}%
\label{fig:4}
\end{figure}

\begin{figure}[h]
\begin{center}
\includegraphics[width=8.5cm]{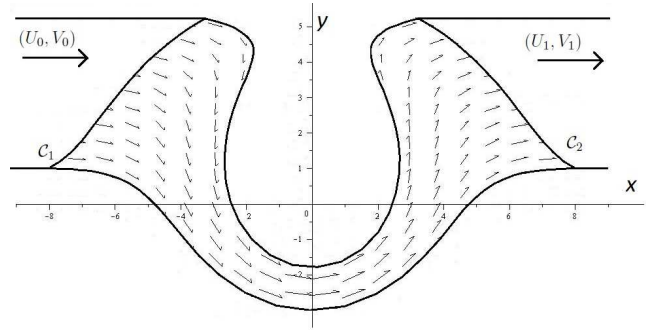}
\end{center}
\caption{Extrusion die corresponding to the solution (\ref{eq:77}), (\ref{eq:109}).}%
\label{fig:5}
\end{figure}

Now, we consider the case where $c_1=0$ in (\ref{eq:73}). Then the solutions for $\theta$ and
$\sigma$ are given by (\ref{eq:77}). We still suppose that the solution for $u$ and $v$ is in the form
(\ref{eq:78}). The procedure is the same as for the previous case until we obtain the conditions
(\ref{eq:82}), (\ref{eq:83}) and (\ref{eq:84}). We must again consider two distinct cases.
\paragraph*{(a.)}We first suppose that the condition (\ref{eq:85}) is satisfied. Then, the functions $f$ and $g$
are defined by
\begin{aleq}\label{eq:99}
f(x,y)=&\int^{\xi(x,y)}(\zeta_2(\xi)+\xi)\zeta'_4(\xi)-\zeta_3(\xi)d\xi-\w_1y+\omega_3,\\
g(x,y)=&\zeta_4(\xi(x,y))+\w_1x+\w_2,\qquad \w_i\in\RR, \ i=1,2,3,
\end{aleq}%
where the $\zeta_i$, $=1,2,3$, are arbitrary functions of one variable and, to simplify the
expression for $f$ and $g$, we have chosen
$\zeta_1(\xi)=\pa{1-\zeta_2(\xi)\xi-\xi^2}\zeta_4(\xi)-\xi\zeta_3(\xi)$. We substitute
(\ref{eq:99}) in equations (\ref{eq:79}) and (\ref{eq:80}) to determine $F$ and $G$. We
conclude that $F(\xi)$ is an arbitrary function, while $G(\xi)$ is expressed as the quadrature
\begin{eqe}\label{eq:100}
G(\xi)=\int\pa{F'(\xi)\xi+\pa{\xi\zeta_2(\xi)-1+\xi^2}\zeta'_4-\xi\zeta_3(\xi)}d\xi.
\end{eqe}%
We finally obtain the solution $u$ and $v$ by introducing (\ref{eq:99}), (\ref{eq:100}) in
(\ref{eq:78}) and, to simplify, by choosing
$$\zeta_3(\xi)=-\pa{\zeta_2(\xi)+\xi}\zeta'_4(\xi),$$%
which gives
\begin{aleq}\label{eq:101}
u(x,y)=&-c_2 y+F'(y/x),\\
v(x,y)=&c_2 x+(y/x)F'(y/x)-F(y/x),
\end{aleq}%
where $F$ is an arbitrary function of one variable. A solution of the system (\ref{eq:1}) consists
of the angle $\theta$ and the pressure $\sigma$ defined by (\ref{eq:77}) with the velocities
defined by (\ref{eq:101}). For example, if we choose the arbitrary function to be an elliptic
function, that is
$$F(\xi)=\operatorname{cn}\pa{\pa{1+\cosh(\arctan(b_2\xi))}^{-1},\varrho},\qquad 0<\rho^2<1,$$%
and we set the parameters as $b_1=4\pi$, $c_2=0$, $\rho=1/2$, then we can trace (see figure \ref{fig:4}) an extrusion die for a feeding speed of
$(U_0,V_0)=(0,-0.94)$ and an extraction speed $(U_1,v_1)=(0,-0.94)$. The curve $C_1$ on the figure \ref{fig:4} delimits the plasticity region at the mouth of the tool, while the $x$-axis does the same for the output of the
tool. This type of tool could be used to undulate a plate. We can shape the tool by varying the
parameters. For example, we can spread the bump by decreasing the parameter $b_1$. Moreover, one
should note that if the modulus $\varrho$ of the elliptic function is such that $0\leq
\varrho^2\leq 1$, then the solution has one purely real and one purely imaginary period. For a real
argument $\chi$, we have the relations
$$-1\leq \operatorname{cn}(\chi,\varrho)\leq 1.$$%
%################################################ figure 4 #######################################
\paragraph*{(b).}Suppose that $\zeta_2(\xi)$ is defined by (\ref{eq:89}) and for simplification we
choose in particular
\begin{eqe}\label{eq:102}
\zeta_3(\xi)=\frac{\zeta_1(\xi)}{\xi}.
\end{eqe}%
Applying the mixed derivatives compatibility condition of $f$ to the equations (\ref{eq:82}),
(\ref{eq:83}), we get the following ODE for the function $g$:
\begin{eqe}\label{eq:103}
g_{xx}-g_{yy}-2\cot(2J(\xi))g_{xy}-\frac{4\xi J'(\xi)g_y}{x\sin^2(2J(\xi))}=0.
\end{eqe}%
By the change of variable
\begin{eqe}\label{eq:104}
\xi(x,y)=y/x,\qquad \eta(x,y)=x^2+y^2,
\end{eqe}%
we reduce the PDE (\ref{eq:103}) in term of $x$ and $y$, to the much simpler PDE in term of $\xi$
and $\eta$,
\begin{eqe}\label{eq:105}
g_{\xi \eta}+\frac{\xi g_\eta}{\xi^2+1}=0,
\end{eqe}%
which has the solution
\begin{eqe}\label{eq:106}
g(\xi,\eta)=\zeta_4(\xi)+\frac{\zeta_5(\eta)}{\sqrt{\xi^2+1}},
\end{eqe}%
where $\zeta_4$ and $\zeta_5$ are arbitrary functions of one variable. Then, we find the solution
for $f$ by integration of the PDE (\ref{eq:82}), (\ref{eq:83}), with $\zeta_3$ given by
(\ref{eq:102}),
\begin{eqe}\label{eq:107}
f(x,y)=-\int^{\xi(x,y)}\frac{\zeta_1(\xi)-\zeta'_4(\xi)}{\xi}d\xi-\frac{y\zeta_5(\eta(x,y))}{\sqrt{\eta(x,y)}}+c_2.
\end{eqe}%
By the substitution of (\ref{eq:106}) and (\ref{eq:107}) in the equations (\ref{eq:79}),
(\ref{eq:80}), we find that $F$ is an arbitrary function of one variable and $G$ is defined by
\begin{eqe}\label{eq:108}
G(\xi(x,y))=\int^{\xi(x,y)}\pa{-\zeta_1(\xi)+F'(\xi)\xi}d\xi+c_3.
\end{eqe}%
So, we introduce (\ref{eq:106}), (\ref{eq:107}) and (\ref{eq:108}) in (\ref{eq:78}) and after an
appropriate redefinition of $\zeta_1$, $\zeta_4$ and $\zeta_5$, the solution of (\ref{eq:1}.c),
(\ref{eq:1}.d) is provided by
\begin{eqe}\label{eq:109}
u(x,y)=K'(y/x)-y H(x^2+y^2)+c_2,\qquad v(x,y)=-K'(y/x)+\xi K(\xi)+x H(x^2+y^2)+c_1,
\end{eqe}%
where $H$, $K$ are arbitrary functions of one variable. The velocities (\ref{eq:109}), together with
the angle and pressure defined by (\ref{eq:77}), solve the system (\ref{eq:1}). For example, a tool
corresponding to the solution (\ref{eq:109}) with $H(\eta)=2\exp(-0.1\eta)$, $K(\xi)=\xi$ and for
feeding and extraction speed given respectively by $(U_0,v_0)=(1.05,0)$ and $(U_1,V_1)=(1.05,0)$. It is shown in figure
\ref{fig:5}. The plasticity region limits correspond to the curves $C_1$ and $C_2$. This tool is
symmetric under the reflection $x\mapsto -x$. Moreover, the top contour of the tool almost makes a
complete loop, and this lets one suppose that we could make a ring in a material by extrusion.
%################################### multiplicative separation when c_1<>0 #####################################################
\subsubsection{Multiplicative separation for the velocities $u$ and $v$ when $c_1\neq 0$.}
Consider the solutions of (\ref{eq:1}.a), (\ref{eq:1}.b), given by the angle $\theta(x,y)=J(y/x)$
with $J$ defined by (\ref{eq:74}) and the pressure $\sigma$ defined by (\ref{eq:75}). We require that
the solutions for $u$ and $v$ be in the multiplicative separated form
\begin{eqe}\label{eq:ms:1}
u(x,y)=f(x,y)F(\xi(x,y)),\quad v(x,y)=g(x,y)G(\xi(x,y)),
\end{eqe}%
where $\xi(x,y)=y/x$ and $f$, $g$, $F$, $G$ are to be determined. The Ansatz (\ref{eq:ms:1}) on the
velocities brings the system (\ref{eq:1}.c), (\ref{eq:1}.d) to the form
\begin{aleq}\label{eq:ms:2}
&\crochet{f_yF+x^{-1}fF'(\xi)+g_xG(\xi)-x^{-1}\xi gG'(\xi)}\sin(2J(\xi))\\
&+\crochet{f_xF(\xi)-x^{-1}\xi
fF'-g_yG-x^{-1}gG'}\cos(2J(\xi))=0,\\
&\crochet{f_xF(\xi)+g_yG(\xi)}-\xi fF'(\xi)+gG'(\xi)=0.
\end{aleq}%
To reduce the PDE system (\ref{eq:ms:2}) to an ODE system involving $F$ and $G$ in terms of $\xi$,
we act with the operator $(1/2)(D_1+D_2)$, defined by (\ref{eq:2}), and annihilate the function of $\xi$
present in (\ref{eq:ms:2}). This leads to conditions on $f$ and $g$ that do not involve $F$, $G$
and their derivatives. There are three cases to consider, that is
\begin{aleq}\label{eq:ms:3}
&(a)\quad (1/2)(D_1+D_2)\pa{g_y/g}\neq0,\qquad (D_1+D_2)\pa{\frac{(1/2)(D_1+D_2)(f/g)}{(D_1+D_2)(xg_y/g)}}\neq0,\\
&(b)\quad (1/2)(D_1+D_2)(x g_y/g)=0,\\
&(c)\quad (1/2)(D_1+D_2)\pa{xg_y/g}\neq0,\qquad (D_1+D_2)\pa{\frac{(1/2)(D_1+D_2)(f/g)}{(D_1+D_2)(xg_y/g)}}=0.
\end{aleq}%
In this paper, we present the details for the cases (a) and (b).
\paragraph*{(a).}We first suppose that the condition (\ref{eq:ms:3}.a) is satisfied. In this case,
the function $f$ must be a solution of the PDE system
\begin{aleq}\label{eq:ms:4}
f_x=&-x^{-1}\xi\zeta_1(\xi)f+\zeta_2(\xi)g_y-x^{-1}\pa{\zeta_2(\xi)\zeta_1'(\xi)-\zeta_2'(\xi)}g,\\
f_y=&x^{-1}\zeta_1(\xi)f+\zeta_2(\xi)g_x-2\zeta_2(\xi)\cot(2J(\xi))g_y\\
&+\pa{\zeta_2(\xi)\zeta_1'(\xi)-\zeta_2'(\xi)}\frac{\xi\sin(2J(\xi))+2\cos(2J(\xi))}{x\sin(2J(\xi))}g,
\end{aleq}%
where $\zeta_1$, $\zeta_2$ are two arbitrary functions of one variable. For the system
(\ref{eq:ms:4}) to be compatible, the function $g$ must satisfy the PDE
\begin{aleq}\label{eq:ms:5}
&g_{xx}-2\cot(2J(\xi))g_{xy}-g_{yy}+2x^{-1}\zeta_2(\xi)^{-1}\pa{\xi+\cot(2J(\xi))}\zeta_3(\xi)g_x\\
&+2x^{-1}\crochet{\pa{\xi-\cot(2J(\xi))}\zeta_3(\xi)-2\xi\sin^{-2}(J(\xi))J'(\xi)}g_y\\
&+x^{-2}\zeta_2(\xi)^{-1}\Big([4\xi
J'(\xi)\sin^{-2}(2J(\xi))+\pa{\xi^2-1+2\xi\cot(2J(\xi))}\zeta_1'(\xi)\\
&-2\xi-2\cot(2J(\xi))]\zeta_3(\xi)-\pa{\xi^2-1+2\xi\cot(2J(\xi))}\zeta_3'(\xi)\Big) g=0,
\end{aleq}%
where we used the notation $\zeta_3(\xi)=\zeta_2(\xi)\zeta_1'(\xi)-\zeta_2'(\xi)$ to shorten the
expression. The equation (\ref{eq:ms:5}) is difficult to solve for arbitrary $\zeta_1$, $\zeta_2$,
but if we make the particular choice
\begin{eqe}\label{eq:ms:6}
\zeta_2(\xi)=\omega_2\exp(\zeta_1(\xi)),\qquad \omega_2\in \RR,
\end{eqe}%
then the PDE (\ref{eq:ms:5}) reduces to
\begin{eqe}\label{eq:ms:7}
g_{xx}-2\cot(2J(\xi))g_{xy}-g_{yy}-4x^{-1}\xi J'(\xi)\sin^{-2}(J(\xi))g_y=0,
\end{eqe}%
which is solved by the function
\begin{eqe}\label{eq:ms:8}
g(x,y)=\omega_3 x+\omega_4\cos(2J(y/x))+\omega_5,\quad \omega_3,\omega_4,\omega_5\in \RR.
\end{eqe}%
With $g$ given by (\ref{eq:ms:8}) and $\zeta_2(\xi)$ by (\ref{eq:ms:6}), the system (\ref{eq:ms:4})
is compatible. Consequently,  $f$ is expressed in term of a quadrature. We find
\begin{aleq}\label{eq:ms:9}
f(x,y)=&2\exp(\zeta_1(y/x))\omega_2\omega_4\int^{y/x}\xi^{-1}\sin(2J(\xi))J'(\xi)d\xi\\
&+\exp(\zeta_1(y/x))\pa{\omega_2\omega_3+2\omega_2\omega_4c_1\ln(y)+\omega_6}
\end{aleq}%
With $f$ given by (\ref{eq:ms:9}) and $g$ by (\ref{eq:ms:8}), the solutions for $F$ and $G$ of the
system (\ref{eq:ms:2}) are
\begin{eqe}\label{eq:ms:10}
F(\xi)=\omega_1\exp(-\zeta_1(\xi)),\qquad G(\xi)=\omega_2.
\end{eqe}%
By introducing (\ref{eq:ms:8}), (\ref{eq:ms:9}), (\ref{eq:ms:10}) in (\ref{eq:ms:1}) and redefining
the free parameters $\omega_i$, $i=1,2,3,4$, the solution of (\ref{eq:1}.c), (\ref{eq:1}.d) for
the velocities $u$ and $v$ is
\begin{aleq}\label{eq:ms:11}
&u(x,y)=2c_4\int^{y/x}\xi^{-1}\sin(2J(\xi))J'(\xi)d\xi+c_5y+2c_1c_4\ln(y)+c_6,\\
&v(x,y)=-c_5x-c_4\cos(2J(y/x))+c_7,
\end{aleq}%
where the $c_i$, $i=1,\ldots,7$, are integration constants and $J$ is defined by (\ref{eq:74}). So,
we have a solution of the system (\ref{eq:1}) composed of the angle $\theta$ in the form
(\ref{eq:formSolProp}) with $J$ given implicitly by (\ref{eq:74}) together with the pressure
$\sigma$ (\ref{eq:75}) and the velocities (\ref{eq:ms:11}).
\paragraph*{(b).} Suppose now that the condition(\ref{eq:ms:3}.c) is
satisfied. In this case, the solution $g$ takes the form
\begin{eqe}\label{eq:ms:12}
g(x,y)=h_1(x)\zeta_1(\xi(x,y))
\end{eqe}%
and the function $f$ must be a solution of
\begin{aleq}\label{eq:ms:13}
f_x=&x^{-1}\xi\zeta_2'(\xi)f+x^{-1}\zeta_3(\xi)h_1(x),\\
f_y=&-2x^{-1}h_1(x)\zeta_3(\xi)\cot(2J(\xi))+e^{-\zeta_2(\xi)}\pa{\int^{y/x}e^{\zeta_2(\xi)}\zeta_3(\xi)
d\xi-\omega_2}h_1'(x)\\
&-x^{-1}\crochet{\zeta_2'(\xi)+\xi\zeta_3(\xi)}.
\end{aleq}%
We omit, due to its complexity, the expression of the compatibility condition on the mixed
derivative of $f$ relative to $x$ and $y$. Nevertheless, making the specific choice
\begin{eqe}\label{eq:ms:14}
h_1(x)=1,\qquad
\zeta_2(\xi)=\ln\pa{\frac{\sin(2J(\xi))}{(\xi^2-1)\sin(2J(\xi))+2\xi\cos(2J(\xi))}}-\ln(\zeta_3(\xi)),
\end{eqe}%
the system (\ref{eq:ms:13}) turns out to be compatible and the solution of (\ref{eq:ms:13}) is
\begin{aleq}\label{eq:ms:15}
f(x,y)=&\zeta_3(y/x)\bigg(c_1^{-1}\pa{1-(y/x)^2-2(y/x)\cot(2J(y/x))}\int^{y/x}
\xi^{-1}\sin(2J(\xi))J'(\xi) d\xi \\
&+\crochet{1-(y/x)^2-2(y/x)\cot(2J(y/x))} (\ln y-\omega_4) \bigg).
\end{aleq}%
The substitution of (\ref{eq:ms:15}), (\ref{eq:ms:12}) and (\ref{eq:ms:14}) in the system
(\ref{eq:ms:2}) results in an ODE system for $F$ and $G$, omitted due to its complexity, which has
the solution
\begin{aleq}\label{eq:ms:16}
&F(\xi)=\frac{\omega_1\sin(2J(\xi))}{\zeta_3(\xi)\pa{(\xi^2-1)\sin(2J(\xi))+2\xi\cos(J(\xi))}},\\
&G(\xi)=\zeta_1(\xi)^{-1}\pa{\omega_1\cos(2J(\xi))-\omega_2}.
\end{aleq}%
We finally obtain a solution for the system (\ref{eq:1}.c), (\ref{eq:1}.d) by the substitution of
(\ref{eq:ms:15}) and (\ref{eq:ms:16}), with $h_1(x)$, $\zeta_2(\xi)$, defined by (\ref{eq:ms:14}),
in (\ref{eq:ms:1}). This leads to
\begin{aleq}\label{eq:ms:17}
&u(x,y)=-c_4\ln(y)-c_1^{-1}c_4\int^{y/x}\xi^{-1}\sin(2J(\xi))J'(\xi)d\xi+c_5,\\
&v(x,y)=(1/2)c_1^{-1}\omega_1\cos(2J(y/x))+c_6,\qquad c_1,c_4,c_5,c_6 \in \RR.
\end{aleq}%
So, the system (\ref{eq:1}) is solved by the angle $\theta$ in the form (\ref{eq:formSolProp}) with
$J$ implicitly defined by (\ref{eq:74}), together with the pressure $\sigma$ (\ref{eq:75}) and the
velocities (\ref{eq:ms:17}).
%#################################  multiplicative separation when c_1=0 #################################################
\subsubsection{Multiplicative separation for the velocities $u$ and $v$ when $c_1=0$.}
Consider now the case where $c_1=0$ in (\ref{eq:73}) so the solution of (\ref{eq:1}.a),
(\ref{eq:1}.b), for $\theta$ and $\sigma$ is (\ref{eq:77}). We suppose that the velocities $u$ and
$v$ are in the form (\ref{eq:ms:1}). Introducing this form for the velocities and $\theta$ defined
by (\ref{eq:77}) in the equations (\ref{eq:1}.c), (\ref{eq:1}.d), leads to the system
(\ref{eq:ms:2}), which reduces to an ODE system for $F$ and $G$ if the functions $f$ and $g$ satisfy
the condition (\ref{eq:ms:3}). The three different constraints (\ref{eq:ms:3}) must be considered
separately.
\paragraph*{(a).}In the first case, where we consider that the conditions (\ref{eq:ms:3}.a) are satisfied, the
functions $f$ and $g$ must satisfy the system (\ref{eq:ms:4}), (\ref{eq:ms:5}). Changing the
variables $(x,y)$ to the new variables $(\xi, \eta)$ defined by
\begin{eqe}\label{eq:ms:18.a}
\xi(x,y)=y/x,\qquad \eta(x,y)=x^2+y^2,
\end{eqe}%
 and considering $\theta$ given by (\ref{eq:77}), the system
(\ref{eq:ms:4}),(\ref{eq:ms:5}), becomes
\begin{aleq}\label{eq:ms:18}
&f_{\xi}=-\zeta_1(\xi)f-\xi^{-1}\zeta_2g_\xi-\xi^{-1}\pa{\zeta_2'(\xi)-\zeta_1(\xi)\zeta_2(\xi)}g,\\
&f_\eta=\xi\zeta_2(\xi)g_\eta,
\end{aleq}%
\begin{eqe}\label{eq:ms:19}
\frac{g_{\xi\eta}}{g_\eta}+\frac{\zeta_2'(\xi)}{\zeta_2(\xi)}+\zeta_1(\xi)+\xi(\xi^2+1)^{-1}=0,
\end{eqe}%
where  $\zeta_1(\xi)$, $\zeta_2(\xi)$ are arbitrary functions of one variable. The solution of the
system (\ref{eq:ms:18}), (\ref{eq:ms:19}), for $f$ and $g$ as functions of $\xi$ and $\eta$ is
\begin{aleq}\label{eq:ms:20}
f(\xi,\eta)=&\xi\zeta_2(\xi)\pa{K(\xi)+\zeta_2(\xi)^{-1}(1+\xi^2)^{-1/2}\pa{H(\eta)e^{-\int\zeta_1(\xi)d\xi}}}\\
&-e^{-\int\zeta_1(\xi)d\xi}\Bigg(\int\xi^{-1}(\xi^2+1)\big[\zeta_2(\xi)+\pa{\zeta_1(\xi)+\xi(\xi^2+1)^{-1}\zeta_2(\xi)}H(\xi)\\
&+\zeta_2(\xi)K'(\xi)\big]d\xi-\omega_1\Bigg),\\
g(\xi,\eta)=&K(\xi)+\pa{e^{\int\zeta_1(\xi)d\xi}\zeta_2(\xi)\sqrt{\xi^2+1}},
\end{aleq}%
where the functions $K(\xi)$ and $H(\eta)$ are arbitrary functions of one variable. Introducing $f$
and $g$ expressed in the initial variables $x$, $y$, by the substitution of (\ref{eq:ms:18.a}) in
(\ref{eq:ms:12}), the system (\ref{eq:ms:2}) is reduced to an ODE system for the functions $F$ and
$G$ in term of $\xi$ which has the solution
\begin{eqe}\label{eq:ms:21}
F(\xi)=\omega_1e^{\int\zeta_1(\xi)d\xi},\qquad G(\xi)=-\omega_1\zeta_2 (\xi)e^{\int
\zeta_1(\xi)d\xi}.
\end{eqe}%
Finally, redefining
$$\zeta_2(\xi)=-(\omega_1K(\xi))^{-1}\exp\pa{\int\zeta_1(\xi) d\xi}Q(\xi),\qquad H(\eta)=\omega_1^{-1}\sqrt{\eta}P(\eta)$$%
where  $Q(\xi)$, $P(\eta)$ are arbitrary functions, and doing the substitution of (\ref{eq:ms:20})
and (\ref{eq:ms:21}) in (\ref{eq:ms:1}), we obtain the solution of (\ref{eq:1}.c), (\ref{eq:1}.d),
for velocities $u$ and $v$ given by
\begin{aleq}\label{eq:ms:22}
&u(x,y)=y P(x^2+y^2)-x^{-1}y Q(y/x)\int^{y/x}\xi^{-1}\pa{(\xi^2+1)Q'(\xi)+\xi Q(\xi)} d\xi+c_2,\\
&v(x,y)=Q(y/x)-x P(x^2+y^2),
\end{aleq}%
This solution for velocities together with $\theta$ and $\sigma$ defined by (\ref{eq:77}) solves
the initial system (\ref{eq:1}).
%######### case (b) ###################
\paragraph*{(b).}Consider now that the functions $f$ and $g$ satisfy
the constraint (\ref{eq:ms:3}.b). Then they take the form
\begin{eqe}\label{eq:ms:22:1}
f(x,y)=\zeta_1(y/x),\qquad g(x,y)=\zeta_2(y/x),
\end{eqe}%
where $\zeta_1$ and $\zeta_2$ are arbitrary functions of one variable. Since the functions $f$ and
$g$ depend only on the symmetry variable and the velocities $u$ and $v$ have the form
(\ref{eq:ms:1}), it is equivalent to consider that
\begin{eqe}\label{eq:ms:22:2}
u(x,y)=F(\xi(x,y)),\qquad v(x,y)=G(\xi(x,y)).
\end{eqe}%
If we suppose that $u$ and $v$ are in the form (\ref{eq:ms:22:2}), then the solution of
(\ref{eq:1}) consists of the angle $\theta$ and the pressure $\sigma$ given by (\ref{eq:77})
together with the velocities
\begin{eqe}\label{eq:ms:22:3}
u(x,y)=F(y/x),\qquad v(x,y)=\int^{y/x}\xi F'(\xi)d\xi + c_3,
\end{eqe}
where $F$ is an arbitrary function of one variable.
%######### case (c) ###################
\paragraph*{(c).}The third case to consider is when $f$ and $g$ obey
the conditions (\ref{eq:ms:3}.c), so they take the form
\begin{eqe}\label{eq:ms:23}
f(x,y)=\omega_3x^{1+\omega_2}(y/x)^{(1+\omega_2)/2}(x^{-1}(x^2+y^2))^{\omega_2/4},\qquad
g(x,y)=x^{1+\omega_2}\zeta_1(y/x),
\end{eqe}%
where $\zeta_1(\xi)$ is an arbitrary function of one variable. Then we introduce (\ref{eq:ms:23})
in (\ref{eq:ms:2}) and solve for $F$ and $G$. The solution is
\begin{eqe}\label{eq:ms:24}
F(\xi)=\omega_1\zeta_1^{-1}(\xi)\xi^{(1-\omega_2)/2}(\xi^2+1)^{\omega_2/4},\qquad
G(\xi)=(\xi^2+1)^{\omega_2/2}.
\end{eqe}%
Finally, substitution of $f$, $g$, $F$, $G$, given by (\ref{eq:ms:23}) and (\ref{eq:ms:24}), in
(\ref{eq:ms:2}) gives, after redefining the parameters $\omega_i$ in a convenient way, the solution
for $u$ and $v$ of the equations (\ref{eq:1}.c), (\ref{eq:1}.d),
\begin{eqe}\label{eq:ms:25}
u(x,y)=c_3y(x^2+y^2)^{\omega_2/2},\qquad v(x,y)=-c_4x(x^2+y^2)^{\omega_2/2},
\end{eqe}%
where $c_3$, $c_4$ are integration constants. The velocities $u$ and $v$ together with the angle
$\theta$ and the pressure $\sigma$ given by (\ref{eq:77}) constitute a solution for the system
(\ref{eq:1}). This solution is just a subcase of the previous one corresponding to the condition
(\ref{eq:ms:3}.a) and the choice $Q(\xi)=0$, $P(\eta)=\eta^{\omega_2/2}$ in (\ref{eq:ms:22}).
%############################################ section #####################################################
\section{Final remarks.}
In this paper, we have obtained the infinitesimal generators which generate the Lie algebra of symmetries for the system (\ref{eq:1}) describing a planar flow of an ideal plastic material. The existence of generators $P_1$ to $P_5$,
$D_1,D_2$, $L$, $B_2$, given in equation (\ref{eq:2}), together with generator $X_1$ of an infinite subalgebra, defined by (\ref{eq:2b}), were already known in the literature \cite{AnninBytevSenashov:1985}. However, we have shown that the symmetry group is completed by the addition of generators $B_1$, $B_3$, $B_4$, $B_5$, $B_6$, $K$, defined by (\ref{eq:2}) and by $X_2$ given to equation (\ref{eq:2b}), which generate an infinite-dimensional subalgebra. We have seen that it is possible to include the generator $K$ in the basis of a finite-dimensional Lie subalgebra only if no generator $B_3$ to $B_6$ and $P_1$ to $P_4$ appears. For this reason, we consider separately the case of the subalgebra $\mathcal{L}$ which excludes $K$ and that of subalgebra $\mathcal{S}$ including $K$, defined by equations (\ref{eq:defAlgL}) and (\ref{eq:defAlgS}) respectively. For each of these two subalgebras, we have performed a classification of the subalgebras into conjugation classes under the action of the symmetry group using the methods described in \cite{PateraWinter:1,PateraWinter:2} (see Section \ref{sec:2} and the appendix). This classification is an important tool in the analysis of invariant and partially invariant solutions. A classification of the symmetry subalgebras of (\ref{eq:1}) has been performed in the past \cite{AnninBytevSenashov:1985} for one-dimensional subalgebras. However, the classification presented here is more complete in the sense that it includes new infinitesimal generators and a classification of two-dimensional subalgebras, which can be used to obtain partially invariant solutions. In section \ref{sec:3}, we have performed (as an example) symmetry reductions corresponding to one-dimensional subalgebras represented by newly found generators. For the first reduction, we have used the generator $B_1$ and for the second reduction the generator $K$, both defined in the list of generators (\ref{eq:2}). The symmetry reduction, using the invariants of $B$, leads to a new solution (see (\ref{ex1:eq:6}), (\ref{ex1:eq:7})) defined in terms of an arbitrary function of $\xi=xu+yv$ and where the velocity fields are implicitly defined. For a particular choice of the arbitrary function, we have traced (in Figure \ref{fig:1}) the shape of an extrusion die corresponding to this solution. The obtained solution has the particularity that the kinetic energy is constant along the flow. A similar analysis has been performed for the generator $K$ in order to obtain a partially invariant solution. For this solution, the invariants given in equation (\ref{ex2:eq:1}) were used to add constraints which allow us to obtain a solution more easily. These considerations were illustrated by finding a particular solution of system (\ref{eq:1}) defined by equations (\ref{ex2:eq:4}), (\ref{ex2:eq:5}) et (\ref{ex2:eq:7}). An example of a velocity vector field and an example of an extrusion die have been traced respectively in figures \ref{fig:2} and \ref{fig:3}. It should be noted that, to  the vector field in Figure \ref{fig:2} (\ie for the corresponding parameter values), we can associate a large family of extrusion dies, of which the one in Figure \ref{fig:3} is a particular choice. The contours have to be chosen in such a way that they correspond to flow lines to the extent that it is possible to trace two curves linking them which satisfy equation (\ref{eq:edoLimPlas}). These curves are the limit of the plasticity region.
\paragraph{}An interesting observation concerning the generator $K$ is that if we take the commutator of $K$ with the generators $\ac{P_1,P_2,P_3,P_4}$, we obtain the generators $\ac{B_3,B_4,B_5,B_6}$. Repeating the procedure with generators $\ac{B_3,B_4,B_5,B_6}$, we generate four new generators, and so on. An interesting fact is that, at each step, the new obtained generators can complete the base of the subalgebra $\mathcal{L}_i$ generated at the previous stage in order to form a new higher-dimensional (but still finite-dimensional) subalgebra (ensured by excluding $K$ from the base). Consequently, it is always possible to enlarge a finite-dimensional subalgebra $\mathcal{L}_i$ that excludes $K$ by increasing its base with the result of commutators $\crochet{K,Z}$ with $Z\in \mathcal{L}_i$. This gives us a chain of finite-subalgebras of the form
$\mathcal{L}\subset\mathcal{L}_1\subset\mathcal{L}_2\subset\ldots\subset\mathcal{L}_i\subset\ldots$.
This subject will be addressed in future works.
\section*{Acknowledgments}The author is greatly indebted to professor A.M. Grundland (Centre de Recherche
Math\'ematiques, Universit\'e de Montr\'eal) for several
valuable and interesting discussions on the topic of this work. This
work was supported by research grants from NSERC of Canada.
\clearpage
\bibliographystyle{unsrt}% plain,unsrt,alpha

%########
\newpage
\section*{Appendix. List of two-dimensional representative subalgebras of $\mathcal{L}$}\label{sec:results}
%
% #######################################   TABLE     #######################
%
\vspace{-5mm}
\begin{table}[h]
\begin{center}
\fontsize{10}{10} \selectfont
\begin{tabular}{|l | l |}
\hline
$\mathcal{L}_{2,1}=\ac{B_1,D_2}$ & $\mathcal{L}_{2,23}=\ac{B_1+\epsilon D_1+\delta P_5,L+aP_5}$\\
$\mathcal{L}_{2,2}=\ac{B_1,D_2+\lambda L+a_1 D_1+a_2 P_5}$ & $\mathcal{L}_{2,24}=\ac{B_1+\epsilon D_1+a P_5,P_1}$\\
$\mathcal{L}_{2,3}=\ac{B_1,D_2+\lambda D_1}$ & $\mathcal{L}_{2,25}=\ac{B_1+\epsilon B_5+a B_6,B_3}$\\
$\mathcal{L}_{2,4}=\ac{B_1,D_2+3D_1+\delta P_5}$ & $\mathcal{L}_{2,26}=\ac{B_1+\epsilon_1 B_5+a_1 B_6,B_3+a_2 P_1+a_3 P_2+\epsilon_2 P_4}$\\
$\mathcal{L}_{2,5}=\ac{B_1,D_2-D_1+\delta P_5}$ & $\begin{aligned}\mathcal{L}_{2,27}=&\{B_1+\epsilon_1 B_5+a_1 B_6+a_2P_4+\epsilon_2 P_5,\\
&B_3+a_3 P_1+a_4P_2+a_4 P_4\}\end{aligned}$\\
$\mathcal{L}_{2,6}=\ac{B_1, D_2+D_1+\delta P_5}$ & $\mathcal{L}_{2,28}=\ac{B_1+\epsilon B_5+a_1 P_4,D_2+3D_1+a_2 P_5}$\\
$\mathcal{L}_{2,7}=\ac{B_1,D_2+\delta P_5}$ & $\mathcal{L}_{2,29}=\ac{B_1+\epsilon B_5+a_1 P_4+a_2 P_5,P_1}$\\
$\mathcal{L}_{2,8}=\ac{B_1,L+a_1 D_1+a_2 P_5}$ & $\mathcal{L}_{2,30}=\ac{B_1+\epsilon P_3, P_5}$\\
$\mathcal{L}_{2,9}=\ac{B_1,D_1+aP_5}$ & $\mathcal{L}_{2,31}=\ac{B_1+\epsilon P_3+a_1 P_4, B_3+a_2 P_2}$\\
$\mathcal{L}_{2,10}=\ac{B_1,B_3}$ & $\mathcal{L}_{2,32}=\ac{B_1+\epsilon_1 P_3+a_1 P_4+\epsilon_2 P_5,B_3+a_2 P_2}$\\
$\mathcal{L}_{2,11}=\ac{B_1,B_3+aP_2+\epsilon P_4}$ & $\mathcal{L}_{2,33}=\ac{B_1+\epsilon_1 P_3+a_1 P_4+a_2 P_5,B_3+a_3 P_2+\epsilon_2 P_4}$\\
$\mathcal{L}_{2,12}=\ac{B_1,B_3+\epsilon_1 P_4+\epsilon_2 P_5}$ & $\mathcal{L}_{2,34}=\ac{B_1+\epsilon_1 P_3+a_1P_4+a_2 P_5, B_3+\epsilon_2 P_4+\delta P_5}$\\
$\mathcal{L}_{2,13}=\ac{B_1,B_3+\epsilon P_5}$ & $\mathcal{L}_{2,35}=\ac{B_1+\epsilon P_3+a_1 P_4+a_2 P_5,B_3+\delta P_5}$\\
$\mathcal{L}_{2,14}=\ac{B_1,P_1}$ & $\mathcal{L}_{2,36}=\ac{B_1+\epsilon P_3+a P_4+\epsilon_2 P_5,P_1}$\\
$\mathcal{L}_{2,15}=\ac{B_1,P_5}$ & $\mathcal{L}_{2,37}=\ac{B_1+\epsilon_1 P_3+a_1 P_4,P_1}$\\
$\mathcal{L}_{2,16}=\ac{B_1+\epsilon L, D_1+a_1 L +a_2 P_5}$ & $\mathcal{L}_{2,38}=\ac{B_1+\epsilon_1 P_4,P_1}$\\
$\mathcal{L}_{2,17}=\ac{B_1+\epsilon L+a D_1,P_5}$ & $\mathcal{L}_{2,39}=\ac{B_1+\epsilon_1 P_4+\epsilon_2 P_5,P_1}$\\
$\mathcal{L}_{2,18}=\ac{B_1+\epsilon L+\delta P_5, D_1+a_1 L+a_2 P_5}$ & $\mathcal{L}_{2,40}=\ac{B_1+\epsilon P_5, L+a_1 D_1+a_2 P_5}$\\
$\mathcal{L}_{2,19}=\ac{B_1+\epsilon D_1,L+aP_5}$ & $\mathcal{L}_{2,41}=\ac{B_1+\epsilon P_5, B_3+a_1 P_2+a_2 P_4}$\\
$\mathcal{L}_{2,20}=\ac{B_1+\epsilon D_1+a_1 P_5,B_3+a_2 P_2}$ & $\mathcal{L}_{2,42}=\ac{B_1+\epsilon P_5, B_3+a P_4+\epsilon_2 P_5}$\\
$\mathcal{L}_{2,21}=\ac{B_1+\epsilon D_1+a_1 P_5, B_3+a_2 P_2+\epsilon P_4}$ & $\mathcal{L}_{2,43}=\ac{B_1+\epsilon P_5,P_1}$\\
$\mathcal{L}_{2,22}=\ac{B_1+\epsilon D_1,P_5}$ & \\
\hline
\end{tabular}
\end{center}
\caption{List of 2-dimensional representative subalgebras of
$\mathcal{L}$ that have a nonzero component $B_1$ but with no $B_2$
component. The parameters are $\epsilon=\pm 1$ and
$a,\lambda,\delta,\in\RR$, where $\delta\neq 0$, $\lambda>0$.}
\end{table}%
\begin{table}[h]
\begin{center}
\fontsize{10}{10} \selectfont
\begin{tabular}{|l | l |}
\hline
$\mathcal{L}_{2,44}=\ac{D_2,L+a_1 D_1 +a_2 P_5}$ & $\mathcal{L}_{2,75}=\ac{D_2+D_1+\epsilon B_5+\delta P_5,P_1}$\\
$\mathcal{L}_{2,45}=\ac{D_2,D_1+a P_5}$ & $\mathcal{L}_{2,76}=\ac{D_2+D_1+\epsilon B_5+\delta P_5,P_u}$\\
$\mathcal{L}_{2,46}=\ac{D_2,B_3+a P_2}$ & $\mathcal{L}_{2,77}=\ac{D_2+D_1+\epsilon B_6, B_5}$\\
$\mathcal{L}_{2,47}=\ac{D_2, B_5 +a P_4}$ & $\mathcal{L}_{2,78}=\ac{D_2+D_1+\epsilon B_6, B_5+\delta P_u + a P_v}$\\
$\mathcal{L}_{2,48}=\ac{D_2,P_1}$ & $\mathcal{L}_{2,79}=\ac{D_2+D_1+\epsilon B_6, B_5+\delta  P_v}$\\
$\mathcal{L}_{2,49}=\ac{D_2,P_3}$ & $\mathcal{L}_{2,80}=\ac{D_2+D_1+\epsilon B_6+\delta P_3, B_5+a_1 P_3+a_2 P_4}$\\
$\mathcal{L}_{2,50}=\ac{D_2,P_5}$ & $\mathcal{L}_{2,81}=\ac{D_2+D_1+\epsilon P_3+a P_4,B_3}$\\
$\mathcal{L}_{2,51}=\ac{D_2+\lambda L, D_1+a L}$ & $\mathcal{L}_{2,82}=\ac{D_2+D_1+\epsilon P_3+a P_4,B_3+\delta P_2}$\\
$\mathcal{L}_{2,52}=\ac{D_2+\lambda L +a D_1,P_5}$ & $\mathcal{L}_{2,83}=\ac{D_2+D_1+\epsilon P_3+a P_4,B_5}$\\
$\mathcal{L}_{2,53}=\ac{D_2-D_1,P_1}$ & $\mathcal{L}_{2,84}=\ac{D_2+D_1+\epsilon P_3+a P_4,B_5+\delta P_4}$\\
$\mathcal{L}_{2,54}=\ac{D_2-D_1,B_3+\epsilon P_5}$ & $\mathcal{L}_{2,85}=\ac{D_2+D_1+\epsilon P_3 + a P_4, B_5+\delta P_5}$\\
$\mathcal{L}_{2,55}=\ac{D_2-D_1+\epsilon B_3,P_1}$ & $\mathcal{L}_{2,86}=\ac{D_2+D_1+\epsilon P_3+a P_4, P_1}$\\
$\mathcal{L}_{2,56}=\ac{D_2-D_1+\epsilon B_3,P_3}$ & $\mathcal{L}_{2,87}=\ac{D_2+D_1+\epsilon P_3+a P_4,P_3}$\\
$\mathcal{L}_{2,57}=\ac{D_2-D_1+\epsilon B_3+a_1P_2+a_2P_5,P_1}$ & $\mathcal{L}_{2,88}=\ac{D_2+D_1+\epsilon P_4,B_3+aP_2}$\\
$\mathcal{L}_{2,58}=\ac{D_2-D_1+\epsilon P_1+a_1 P_2, B_3+a_2 P_2}$ & $\mathcal{L}_{2,89}=\ac{D_2+D_1+\epsilon P_4,B_5+aP_v}$\\
$\mathcal{L}_{2,59}=\ac{D_2-D_1+\epsilon P_1 +a P_2, B_3+\delta P_5}$ & $\mathcal{L}_{2,90}=\ac{D_2+D_1+\epsilon P_4,B_5+\delta P_5}$\\
$\mathcal{L}_{2,60}=\ac{D_2-D_1+\epsilon P_1+a P_2,B_5}$ & $\mathcal{L}_{2,91}=\ac{D_2+D_1+\epsilon P_4,P_1}$\\
$\mathcal{L}_{2,61}=\ac{D_2-D_1+\epsilon P_1+a P_2,B_5+\delta P_4}$ & $\mathcal{L}_{2,92}=\ac{D_2+D_1+\epsilon P_4,P_3}$\\
$\mathcal{L}_{2,62}=\ac{D_2-D_1+\epsilon P_1+a P_2,P_3}$ & $\mathcal{L}_{2,93}=\ac{D_2+D_1+\delta P_5,P_3}$\\
$\mathcal{L}_{2,63}=\ac{D_2-D_1+\epsilon P_1+a P_2,P_5}$ & $\mathcal{L}_{2,94}=\ac{D_2+\lambda D_1, L+aP_5}$\\
$\mathcal{L}_{2,64}=\ac{D_2-D_1+\epsilon  P_2,B_3+a_2 P_2}$ & $\mathcal{L}_{2,95}=\ac{D_2+\lambda D_1, B_3+a P_4}$\\
$\mathcal{L}_{2,65}=\ac{D_2-D_1+\epsilon P_2,B_3+\delta P_5}$ & $\mathcal{L}_{2,96}=\ac{D_2+\lambda D_1, B_5+a P4}$\\
$\mathcal{L}_{2,66}=\ac{D_2-D_1+\epsilon P_2,B_5+a P_4}$ & $\mathcal{L}_{2,97}=\ac{D_2+\lambda D_1, P_3}$\\
$\mathcal{L}_{2,67}=\ac{D_2-D_1+\epsilon P_2,P_3}$ & $\mathcal{L}_{2,98}=\ac{D_2+\lambda D_1,P_5}$\\
$\mathcal{L}_{2,68}=\ac{D_2-D_1+\epsilon P_2,P_5}$ & $\mathcal{L}_{2,99}=\ac{D_2+\lambda D_1+\delta P_5,L+aP_5}$\\
$\mathcal{L}_{2,69}=\ac{D_2-D_1+\delta P_5,P_3}$ & $\mathcal{L}_{2,100}=\ac{D_2+\lambda D_1+aP_5,P_1}$\\
$\mathcal{L}_{2,70}=\ac{D_2+D_1,P_2}$ & $\mathcal{L}_{2,101}=\ac{D_2+\lambda P_5, L+a_1 D_1+a_2 P_5}$\\
$\mathcal{L}_{2,71}=\ac{D_2+D_1, B_5+\epsilon P_5}$ & $\mathcal{L}_{2,102}=\ac{D_2+\lambda P_5, D_1+aP_5}$\\
$\mathcal{L}_{2,72}=\ac{D_2+D_1+\epsilon B_5,P_1}$ & $\mathcal{L}_{2,103}=\ac{D_2+\lambda P_5,P_1}$\\
$\mathcal{L}_{2,73}=\ac{D_2+D_1+\epsilon B_5+\delta P_4,P_1}$ & $\mathcal{L}_{2,104}=\ac{D_2+\lambda P_5,P_3}$\\
$\mathcal{L}_{2,74}=\ac{D_2+D_1+\epsilon B_5+aP_4,P_u}$ & \\
\hline
\end{tabular}
\end{center}
\caption{List of 2-dimensional representative subalgebras of
$\mathcal{L}$ that have a nonzero component $D_2$ but with no
component in $\ac{B_1,B_2}$. The parameters are $\epsilon=\pm
1$ and $a,\lambda,\delta,\in\RR$, where $\delta\neq 0$,
$\lambda>0$.}
\end{table}%
\ \vspace{-5in}\\
\ 
\begin{table}[h]
\begin{center}
\fontsize{10}{10} \selectfont
\begin{tabular}{|l | l |}
\hline
$\mathcal{L}_{2,105}=\ac{B_1-B_2,P_5}$ &$\mathcal{L}_{2,114}=\ac{B_1-B_2+\epsilon_1 L +\delta P_3,P_1-\epsilon_1 P_3+\epsilon_2 P_5}$\\
$\mathcal{L}_{2,106}=\ac{B_1-B_2+\delta L, D_1+a L}$ &$\begin{aligned}\mathcal{L}_{2,115}=\{&B_1-B_2+\epsilon_1 L + aP_3 +\delta P_4,\\
 &\ P_1-\epsilon_1 P_3+\epsilon_2 P_5\}\end{aligned}$\\
$\mathcal{L}_{2,107}=\ac{B_1-B_2+\epsilon_1 L, P_1-\epsilon_1 P_3+\epsilon_2 P_5}$ &$\begin{aligned}\mathcal{L}_{2,116}=\{&B_1-B_2+\epsilon_1 L +a_1 P_3+a_2 P_4+\delta P_5,\\
&\ P_1-\epsilon_1 P_3+\epsilon P_5\}\end{aligned}$\\
$\mathcal{L}_{2,108}=\ac{B_1-B_2+\epsilon L +a D_1, P_1-\epsilon P_3}$ &$\mathcal{L}_{2,117}=\ac{B_1-B_2+\delta D_1, P_5}$\\
$\mathcal{L}_{2,109}=\ac{B_1-B_2+\delta L+a D_1,P_5}$ &$\mathcal{L}_{2,118}=\ac{B_1-B_2+\delta D_1+a_1 P_5, L+a_2 P_5}$\\
$\begin{aligned} \mathcal{L}_{2,110}=\{&B_1-B_2+\epsilon L
+a_1D_1+a_2 P_1+a_2P_2,\\
&\ B_3-\epsilon B_5+a_4P_2+a_5 P_4+a_6P_5\}
\end{aligned}$ &$\mathcal{L}_{2,119}=\ac{B_1-B_2+a_1 P_5, L+a_2 D_1+a_3P_5}$\\
$\begin{aligned}\mathcal{L}_{2,111}=\{&B_1-B_2+\epsilon L +a D_1+a_1 P_1+a_2 P_2,\\
&\ B_4-\epsilon B_6+a_3 P_1+a_4 P_3\}\end{aligned}$ &$\mathcal{L}_{2,120}=\ac{B_1-B_2+a_1 P_5, D_1+a_2 P_5}$\\
$\mathcal{L}_{2,112}=\ac{B_1-B_2+\epsilon L +a D_1 +\delta P_5, P_1-\epsilon P_3}$ &$\mathcal{L}_{2,121}=\ac{L+aD_1,P_5}$\\
$\mathcal{L}_{2,113}=\ac{B_1-B_2+\epsilon_1 L+\epsilon_2 P_1, P_5}$ &$\mathcal{L}_{2,122}=\ac{L+a_1P_5,D_1+a_2 P_5}$\\
\hline
\end{tabular}
\end{center}
\caption{List of 2-dimensional representative subalgebras of
$\mathcal{L}$ that have a nonzero component $B_2$. The parameters are $\epsilon=\pm 1$ and $a,\lambda,\delta,\in\RR$, where
$\delta\neq 0$, $\lambda>0$.}
\end{table}%
\vspace{-50mm}\ \\
\begin{table}[h]
\begin{center}
\fontsize{10}{10} \selectfont
\begin{tabular}{|l | l |}
\hline
$\mathcal{L}_{2,123}=\ac{D_1,B_3}$ & $\mathcal{L}_{2,127}=\ac{D_1,B_3+aP_2+\epsilon P_4}$\\
$\mathcal{L}_{2,124}=\ac{D_1,B_3+\epsilon B_5+a_1 P_1+a_2 P_2+a_3 P_4}$ & $\mathcal{L}_{2,128}=\ac{D_1+a P_5,P_1+\epsilon P_3}$\\
$\mathcal{L}_{2,125}=\ac{D_1,B_3+\epsilon P_3}$ & $\mathcal{L}_{2,129}=\ac{D_1+a P_5,P_3}$\\
$\mathcal{L}_{2,126}=\ac{D_1,B_3+a_1 P_2+\epsilon P_3+a_2 P_4}$ & \\
\hline
\end{tabular}
\end{center}
\caption{List of 2-dimensional representative subalgebras of
$\mathcal{L}$ that do not have components in $\ac{B_1,D_2,B_2,L}$
with a nonzero $D_1$ component. The parameters are
$\epsilon=\pm 1$ and $a,\lambda,\delta,\in\RR$, where $\delta\neq
0$, $\lambda>0$.}
\end{table}%
\begin{landscape}
\begin{table}[t]
\begin{center}
\fontsize{10}{10} \selectfont
\begin{tabular}{|l | l |}
\hline
$\mathcal{L}_{2,130}=\ac{B_3,P_1+a P_2}$ & $\mathcal{L}_{2,163}=\ac{B_3+\epsilon B_5+\delta P_5+a P_4,P_2}$\\
$\mathcal{L}_{2,131}=\ac{B_3+\epsilon_1 B_5,P_1+aP_2+\epsilon_2 P_3}$ & $\mathcal{L}_{2,164}=\ac{B_3+\epsilon B_5+a P_4,P_2}$\\
$\mathcal{L}_{2,132}=\ac{B_3+\epsilon_1 B_5,P_1+\epsilon_2 P_3}$ & $\mathcal{L}_{2,165}=\ac{B_3+\epsilon B_5 +a P_4,P_2}$\\
$\mathcal{L}_{2,133}=\ac{B_3+\epsilon_1 B_5,P_2+\epsilon_2 P_3}$ & $\mathcal{L}_{2,166}=\ac{B_3+\epsilon B_5+\delta P_3+aP_4,P_2+\epsilon_2 P_3}$\\
$\mathcal{L}_{2,134}=\ac{B_3+\epsilon P_3+a P_4, P_1+\delta P_2}$ & $\mathcal{L}_{2,167}=\ac{B_3+\epsilon_1 B_5+\delta P_4,P_2+\epsilon_2 P_3}$\\
$\mathcal{L}_{2,135}=\ac{B_3+\epsilon P_4,P_1+\delta P_2}$ & $\mathcal{L}_{2,168}=\ac{B_3+\epsilon_1 B_5+\delta P_5,P_2+\epsilon_2 P_3}$\\
$\mathcal{L}_{2,136}=\ac{B_3+a_1P_2,P_1}$ & $\mathcal{L}_{2,169}=\ac{B_3+\epsilon B_5+\delta P_2+aP_4,P_3}$\\
$\mathcal{L}_{2,137}=\ac{B_3+a_1P_2+\epsilon P_3+a_2P_4,P_1}$ & $\mathcal{L}_{2,170}=\ac{B_3+\epsilon B_5+aP_4,P_3}$\\
$\mathcal{L}_{2,138}=\ac{B_3+aP_2+\epsilon P_4,P_1}$ & $\mathcal{L}_{2,171}=\ac{B_3+\epsilon B_5+\delta P_5,P_3}$\\
$\mathcal{L}_{2,139}=\ac{B_3+\epsilon P_5,P_1+aP_2}$ & $\mathcal{L}_{2,172}=\ac{B_3+a_1P_1+a_2P_2,B_3+a_3P_1+\epsilon P_3+a_4 P_4}$\\
$\mathcal{L}_{2,140}=\ac{B_3+\delta P_2+a_1 P_4,P_1+\epsilon P_3+a_2 P_4}$ & $\mathcal{L}_{2,173}=\ac{B_3+a_1P_1+a_2P_2,B_4+a_3 P_1+\epsilon P_4}$\\
$\mathcal{L}_{2,141}=\ac{B_3+a_1 P_4,P_1+\epsilon P_3 +a_2 P_4}$ & $\mathcal{L}_{2,174}=\ac{B_3+a_1P_1+a_2P_2,B_4+a_3P_1}$\\
$\mathcal{L}_{2,142}=\ac{B_3+\epsilon P_5,P_1+\epsilon P_3+a P_4}$ & $\mathcal{L}_{2,175}=\ac{B_3+a_1P_1+a_2P_2+\epsilon P_3+a_3 P_4,B_4+a_4 P_1+a_5 P_3+a_6P_4}$\\
$\mathcal{L}_{2,143}=\ac{B_3+\delta P_3+a P_4,P_2+\epsilon P_4}$ & $\mathcal{L}_{2,176}=\ac{B_3+a_1 P_1+a_2 P_2+\epsilon P_4,B_4+a_3 P_1+a_4 P_3+a_5 P_5}$\\
$\mathcal{L}_{2,144}=\ac{B_3+\delta P_4,P_2+\epsilon P_4}$ & $\mathcal{L}_{2,177}=\ac{B_3+a_1P_1+a_2 P_2,B_4+\epsilon B_6 +a_3 P_1+a_4 P_2 +a_5 P_3}$\\
$\mathcal{L}_{2,145}=\ac{B_3+aP_5,P_2+\epsilon P_4}$ & $\mathcal{L}_{2,178}=\ac{B_3+a_1 P_1+a_2 P_2,B_5+a_3 B_6+a_4 P_4}$\\
$\mathcal{L}_{2,146}=\ac{B_3+a_1 P_2, P_3+a_2 P_4}$ & $\mathcal{L}_{2,179}=\ac{B_3+a_1 P_1+a_2 P_2,B_5+a_3 B_6+\epsilon P_x +a_4 P_2+a_5 P_5}$\\
$\mathcal{L}_{2,147}=\ac{B_3+a_1 P_2+\epsilon P_4,P_3+a_2 P_4}$ & $\mathcal{L}_{2,180}=\ac{B_3+a_1 P_x+a_2 P_2,B_5 + a_3 B_6+\epsilon P_2 + a_4 P_4}$\\
$\mathcal{L}_{2,148}=\ac{B_3+\epsilon P_5, P_3+aP_4}$ & $\mathcal{L}_{2,181}=\ac{B_3+a_1 P_1+a_2 P_2+\epsilon P_3+a_3 P_v,B_5+a_4 B_6+a_5 P_1+a_6P_2+a_7 P_4}$\\
$\mathcal{L}_{2,149}=\ac{B_3+a P_2,P_4}$ & $\mathcal{L}_{2,182}=\ac{B_3+a_1P_1+a_2P_2+\epsilon P_4,B_5+a_3 B_6+a_4 P_1+a_5 P_2+a_6 P_4}$\\
$\mathcal{L}_{2,150}=\ac{B_3+a P_2+\epsilon P_3,P_4}$ & $\mathcal{L}_{2,183}=\ac{B_3+a_1 P_1+a_2P_2+\epsilon P_3+a_3 P_4,B_6+a_4 P_1+a_5P_2+a_6P_3}$\\
$\mathcal{L}_{2,151}=\ac{B_3+\epsilon P_5,P_4}$ & $\mathcal{L}_{2,184}=\ac{B_3+a_1P_1+a_2P_2+\epsilon P_4,B_6+a_3 P_1+a_4 P_2 +a_5 P_3}$\\
$\mathcal{L}_{2,152}=\ac{B_3+\epsilon_1 B_5+\delta P_2,P_1+aP_2}$ & $\mathcal{L}_{2,185}=\ac{B_3+a_1 P_1+a_2 P_2,B_6+\epsilon P_1+a_3P_2+a_4P_3}$\\
$\mathcal{L}_{2,153}=\ac{B_3+\epsilon_1 B_5+a_1 P_2+\epsilon_2 P_4,P_1+a_2 P_y}$ & $\mathcal{L}_{2,186}=\ac{B_3+a_1 P_1+a_2 P_2, B_6+\epsilon P_2 + a_4 P_3}$\\
$\mathcal{L}_{2,154}=\ac{B_3+\epsilon B_5+a_1 P_5,P_1+a_2 P_2}$ & $\mathcal{L}_{2,187}=\ac{B_3+a_1 P_1+a_2 P_2,B_6+a_3 P_3}$\\
$\mathcal{L}_{2,155}=\ac{B_3+\epsilon_1 B_5+\delta P_2+a_1 P_3+a_2 P_4,P_1+a_3 P_2+\epsilon_2 P_3}$ & $\mathcal{L}_{2,188}=\ac{B_3+a_1 B_4+a_2 P_1+a_2 P_2+\epsilon P_3+a_3 P_4,B_5+a_4 P_1+a_5P_2+a_6 P_4}$\\
$\mathcal{L}_{2,156}=\ac{B_3+\epsilon_1 B_5 +\delta P_3 +a_1 P_4,P_1+a_2 P_y+\epsilon_2 P_3}$ & $\mathcal{L}_{2,189}=\ac{B_3+a_1B_4+a_2 P_1+a_3 P_2+\epsilon P_4,B_5+a_4 P_1+ a_4 P_2 +a_5 P_4}$\\
$\mathcal{L}_{2,157}=\ac{B_3+\epsilon_1 B_5+\delta P_4,P_1+a_2 P_2 +\epsilon_2 P_u}$ & $\mathcal{L}_{2,190}=\ac{B_3+a_1 B_4+a_2 P_1+a_2P_2,B_5+\epsilon P_1+a P_2+a P_4}$\\
$\mathcal{L}_{2,158}=\ac{B_3+\epsilon_1 B_5+\delta P_5,P_1+aP_2+\epsilon_2 P_3}$ & $\mathcal{L}_{2,191}=\ac{B_3+a_1 B_4+a_2 P_1+a_2 P_2,B_5+\epsilon P_4+a_3 P_4}$\\
$\mathcal{L}_{2,159}=\ac{B_3+\epsilon_1 B_5+\delta P_2+a_1 P_3+a_2 P_4,P_1+\epsilon_2 P_3}$ & $\mathcal{L}_{2,192}=\ac{B_3+a_1 B_4+a_2 P_1+a_3 P_2,B_5+a_4 P_4}$\\
$\mathcal{L}_{2,160}=\ac{B_3+\epsilon_1 B_5+\delta P_3+ a P_v,P_1+\epsilon_2 P_u}$ & $\begin{aligned}\mathcal{L}_{2,193}=\{&B_3+\epsilon B_5+a_1 P_1+a_2 P_2+a_3 P_3+a_4 P_4,\\
&\ B_4+a_5 B_5+a_6 B_6+a_7 P_1+a_8 P_3+a_9 P_4\}\end{aligned}$\\
$\mathcal{L}_{2,161}=\ac{B_3+\epsilon_1 B_5 +\delta P_4,P_1+\epsilon_2 P_3}$ & $\mathcal{L}_{2,194}=\ac{B_3+\epsilon B_5+a_1 P_1+a_2 P_2+a_3 P_3+a_4 P_4, B_5+a_5 B_6+a_6 P_1+a_7 P_2+a_8 P_4}$\\
$\mathcal{L}_{2,162}=\ac{B_3+\epsilon_1 B_5+\delta P_5,P_1+\epsilon_2 P_u}$ & $\mathcal{L}_{2,195}=\ac{B_3+\epsilon B_5+a_1 P_1+a_2 P_2+a_3 P_3+a_4 P_4,B_6+a_5 P_1+a_6 P_2 +a_7 P_3}$\\
\hline
\end{tabular}
\end{center}
\caption{List of 2-dimensional representative subalgebras of
$\mathcal{L}$ that do not have components in $\ac{B_1,D_2,B_2,L,D_1}$
with a nonzero $B_3$ component. The parameters are
$\epsilon=\pm 1$ and $a,\lambda,\delta,\in\RR$, where $\delta\neq
0$, $\lambda>0$.}
\end{table}%
\end{landscape}
\begin{table}[t]
\begin{center}
\fontsize{10}{10} \selectfont
\begin{tabular}{|l | l |}
\hline
$\mathcal{L}_{2,196}=\ac{B_5,B_6}$ & $\mathcal{L}_{2,209}=\ac{B_5+a P_4,P_2+\epsilon P_4}$\\
$\mathcal{L}_{2,197}=\ac{B_5,P_3+aP_4}$ & $\mathcal{L}_{2,210}=\ac{B_5+\delta P_5,P_2+\epsilon P_4}$\\
$\mathcal{L}_{2,198}=\ac{B_5,P_4}$ & $\mathcal{L}_{2,211}=\ac{B_5+\epsilon P_1 +a_1 P_2 +a_2 P_4, P_3+a_3 P_v}$\\
$\mathcal{L}_{2,199}=\ac{B_5+\epsilon P_2+a_1P_4,P_1+a_2 P_2}$ & $\mathcal{L}_{2,212}=\ac{B_5+\epsilon P_2+a_1 P_4, P_3+a_2 P_4}$\\
$\mathcal{L}_{2,200}=\ac{B_5+a_1 P_4,P_1+a_2 P_2}$ & $\mathcal{L}_{2,213}=\ac{B_5+\delta P_4,P_3+a_2 P_4}$\\
$\mathcal{L}_{2,201}=\ac{B_5+\epsilon P_5,P_1+a P_2}$ & $\mathcal{L}_{2,214}=\ac{B_5+\epsilon P_5,P_3+aP_4}$\\
$\mathcal{L}_{2,202}=\ac{B_5+\epsilon_1 P_2+a_1P_4, P_1+a_2P_2+\epsilon_2P_3}$ & $\mathcal{L}_{2,215}=\ac{B_5+\epsilon P_1 +a P_2,P_4}$\\
$\mathcal{L}_{2,203}=\ac{B_5+a_1 P_4, P_1+a_2 P_2+\epsilon P_3}$ & $\mathcal{L}_{2,216}=\ac{B_5+\epsilon P_2,P_4}$\\
$\mathcal{L}_{2,204}=\ac{B_5+\epsilon_1 P_5, P_1+aP_2+\epsilon_2 P_3}$ & $\mathcal{L}_{2,217}=\ac{B_5+\epsilon P_5,P_4}$\\
$\mathcal{L}_{2,205}=\ac{B_5+\epsilon P_1+a P_4,P_2}$ & $\begin{aligned}\mathcal{L}_{2,218}=\{&B_5+a_1 P_3+a_2 P_4,\\
&\  B_6+\epsilon P_1+a_3 P_1+a_3 P_2+a_4 P_3\}\end{aligned}$\\
$\mathcal{L}_{2,206}=\ac{B_5+a P_4, P_2}$ & $\mathcal{L}_{2,219}=\ac{B_5+a_1 P_3+a_2 P_4, B_6+\epsilon P_2+a_3 P_3}$\\
$\mathcal{L}_{2,207}=\ac{B_5+\epsilon P_5, P_2}$ & $\begin{aligned}\mathcal{L}_{2,220}=\{&B_5+\epsilon P_1+a_1P_2+a_2P_3+a_3P_4,\\
&\ B_6+a_4P_1+a_5P_2+a_6P_3\}\end{aligned}$\\
$\mathcal{L}_{2,208}=\ac{B_5+\delta P_1+a P_4, P_2 +\epsilon P_4}$ & $\mathcal{L}_{2,221}=\ac{B_5+\epsilon P_2+a_1 P_3+a_2 P_4, B_6+a_3 P_1+a_4 P_2}$\\
\hline
\end{tabular}
\end{center}
\caption{List of 2-dimensional representative subalgebras of
$\mathcal{L}$ that do not have components in
$\ac{B_1,D_2,B_2,L,D_1,B_3,B_4}$ with a nonzero $B_5$ component. The
parameters are $\epsilon=\pm 1$ and $a,\lambda,\delta,\in\RR$,
where $\delta\neq 0$, $\lambda>0$.}
\end{table}%
\begin{table}[t]
\begin{center}
\fontsize{10}{10} \selectfont
\begin{tabular}{|l | l |}
\hline
$\mathcal{L}_{2,222}=\ac{P_1,P_5}$ & $\mathcal{L}_{2,232}=\ac{P_1+\epsilon P_5,P_4}$\\
$\mathcal{L}_{2,223}=\ac{P_1+aP_2,P_5}$ & $\mathcal{L}_{2,233}=\ac{P_2,P_5}$\\
$\mathcal{L}_{2,224}=\ac{P_1+aP_2+\epsilon_1 P_5,P_2+\epsilon_2 P_5}$ & $\mathcal{L}_{2,234}=\ac{P_2+a_1P_3+a_2P_4,P_5}$\\
$\mathcal{L}_{2,225}=\ac{P_1+\epsilon P_3,P_5}$ & $\mathcal{L}_{2,235}=\ac{P_2+\epsilon P_5,P_3+\epsilon_2 P_5}$\\
$\mathcal{L}_{2,226}=\ac{P_1+\epsilon_1 P_3+\epsilon_2 P_5, P_2+a_1 P_3+a_2P_4+\delta P_5}$ & $\mathcal{L}_{2,236}=\ac{P_2+\epsilon_1 P_5,P_3+\epsilon_2 P_5}$\\
$\mathcal{L}_{2,227}=\ac{P_1+\epsilon_1 P_3+\epsilon_2 P_5,P_v+\delta P_5}$ & $\mathcal{L}_{2,237}=\ac{P_3,P_5}$\\
$\mathcal{L}_{2,228}=\ac{P_1+\epsilon_1 P_5,P_2+\epsilon_2 P_4}$ & $\mathcal{L}_{2,238}=\ac{P_3+aP_4,P_3}$\\
$\mathcal{L}_{2,229}=\ac{P_1+\epsilon P_5,P_2+\delta P_5}$ & $\mathcal{L}_{2,239}=\ac{P_3+\epsilon P_5, P_4+\delta P_5}$\\
$\mathcal{L}_{2,230}=\ac{P_1+\epsilon P_5, P_3+\delta P_4}$ & $\mathcal{L}_{2,240}=\ac{P_4,P_5}$\\
$\mathcal{L}_{2,231}=\ac{P_1+\epsilon_1 P_5, P_3+\epsilon_2 P_5}$ & \\
\hline
\end{tabular}
\end{center}
\caption{List of 2-dimensional representative subalgebras of
$\mathcal{L}$ corresponding to translations with the $P_5$ component.
The parameters are $\epsilon=\pm 1$ and
$a,\lambda,\delta,\in\RR$, where $\delta\neq 0$, $\lambda>0$.}
\end{table}%
\begin{table}[t]
\begin{center}
\fontsize{10}{10} \selectfont
\begin{tabular}{|l | l |}
\hline
$\mathcal{L}_{2,241}=\ac{P_1,P_2}$ & $\mathcal{L}_{2,246}=\ac{P_1+\epsilon P_3, P_2+a_1 P_3+a_2 P_4}$\\
$\mathcal{L}_{2,242}=\ac{P_1,P_2+\epsilon P_4}$ & $\mathcal{L}_{2,247}=\ac{P_1+\epsilon P_3,P_3+a P_4}$\\
$\mathcal{L}_{2,243}=\ac{P_1,P_3+a P_4}$ & $\mathcal{L}_{2,248}=\ac{P_1+\epsilon P_3,P_4}$\\
$\mathcal{L}_{2,244}=\ac{P_1,P_4}$ & $\mathcal{L}_{2,249}=\ac{P_2,P_3}$\\
$\mathcal{L}_{2,245}=\ac{P_1+aP_2,P_3}$ & $\mathcal{L}_{2,250}=\ac{P_3,P_4}$\\
\hline
\end{tabular}
\end{center}
\caption{List of 2-dimensional representative subalgebras of
$\mathcal{L}$ corresponding to translations without the $P_5$ component.
The parameters are $\epsilon=\pm 1$ and
$a,\lambda,\delta,\in\RR$, where $\delta\neq 0$, $\lambda>0$.}
\end{table}%
\end{document}